\documentclass[trackchanges]{aastex701}

\usepackage{amsmath}
\usepackage{mathrsfs}
\usepackage{svg}

\newenvironment{aascases}
{\left\{\begin{array}{ll}}
{\end{array}\right.}

\newcommand{\curlc}{\mathscr{C}}
\newcommand{\pb}{{\overline{p}}}
\newcommand{\epsOoE}{\epsilon \equiv \frac{11}{2(\pb + 1)}\frac{\epsilon_B}{\xi \epsilon_e}}
\newcommand{\xiHP}{\xi \equiv \frac{1-\epsilon_B}{\epsilon_e}}

\newcommand{\shenzhangfactor}{\curlc = 
    \begin{aascases}
        \frac{(p + 2)(p - 1/3)}{p + 2/3}, & \nu_a < \nu_m, \\
        \sqrt{2}(p + 1)
        \frac{
        G\left(\frac{3p + 22}{12}\right)
        G\left(\frac{3p + 2}{12}\right)
        }{
        G\left(\frac{3p + 19}{12}\right)
        G\left(\frac{3p - 1}{12}\right)
        }, & \nu_m < \nu_a,
    \end{aascases}
}

\newcommand{\pbar}{
    \pb\equiv
    \begin{aascases}
        p, & \nu_m<\nu_a \\ 
        2, & \nu_a<\nu_m
    \end{aascases}
}

\newcommand{\etaeq}{
    \eta=
    \begin{aascases}
        \nu_m/\nu_a, & \nu_a < \nu_m \\
        1, & \nu_m < \nu_a.
    \end{aascases}
}

\newcommand{\kappaeq}{
    \kappa=
    \begin{aascases}
        \gamma_e/\gamma_m=(\nu_a/\nu_m)^{\frac{1}{2}}, & \nu_m < \nu_a \\
        1, & \nu_a < \nu_m
    \end{aascases}
}

\newcommand{\chie}{\chi_e \equiv \frac{p-2}{p-1} \epsilon_e \frac{m_p}{m_e}}

\newcommand{\gammaM}{\gamma_m = \mu \chi_e (\Gamma - 1)}

\newcommand{\ReqNdev}{R_{eq, N}^{+\rm{dev}}=\left(\xi\epsilon\gamma_m^{2-\pb}\frac{(\pb+1)\curlc^{\pb+5} c F_p^{\pb+6} d_L^{2(\pb+6)}\eta^{5/3(\pb+5)}}{2^{\pb+2}11\sqrt{3}\pi^{\pb+7}m_e^{\pb+6}\nu_p^{2\pb+13}(1+z)^{3\pb+19}f_A^{\pb+5}f_V}\right)^\frac{1}{2\pb+13}}

\newcommand{\Reqdev}{R_{\rm{eq}}^{+\rm{dev}}=\Gamma\delta_D^{-\frac{\pb+5}{2\pb+13}}R_{\rm{\rm{eq,N}}}^{+\rm{dev}}}

\newcommand{\ReqN}{R_{\rm{\rm{eq,N}}}=\left(\xi\gamma_m^{2-\pb}\frac{(\pb+1)\curlc^{\pb+5} c F_p^{\pb+6} d_L^{2(\pb+6)}\eta^{5/3(\pb+5)}}{2^{\pb+2}11\sqrt{3}\pi^{\pb+7}m_e^{\pb+6}\nu_p^{2\pb+13}(1+z)^{3\pb+19}f_A^{\pb+5}f_V}\right)^\frac{1}{2\pb+13}}

\newcommand{\Req}{R_{\rm{eq}}=\Gamma\delta_D^{-\frac{\pb+5}{2\pb+13}}R_{\rm{\rm{eq,N}}}}

\newcommand{\Etot}{
    \begin{align}
        E_{tot} &= E_e + E_o + E_B
        = \left(1 + \frac{\epsilon_o}{\epsilon_e}\right)E_e + E_B
        = \xi E_e + E_B \\
        &= E_{\rm{eq}}\left(\frac{17}{2\pb + 13}\right)\left(\frac{11}{17}\left(\frac{R}{R_{\rm{eq}}}\right)^{-2(\pb+1)}+\frac{2(\pb+1)}{17}\left(\frac{R}{R_{\rm{eq}}}\right)^{11}\right)
    \end{align}
}

\newcommand{\energyeqNdev}{E_{\rm{\rm{eq,N}}}^{+\rm{dev}}=\left(\xi^{11}\gamma_m^{11(2-\pb)}\frac{17^{2\pb+13}\pi^{\pb+1}\curlc^{3(\pb+1)}c^{4\pb+37}m_e^{\pb+12}F_p^{3\pb+14}d_L^{2(3\pb+14)}\eta^{5(\pb+1)}f_V^{2(\pb+1)}}{2^{7\pb-4}(\pb+1)^{2(\pb+1)}3^5 11^{11} \sqrt{3}q_e^{2(2\pb+13)}\nu_p^{2\pb+13}(1+z)^{5\pb+27}f_A^{3(\pb+1)}}\right)^\frac{1}{2\pb+13}\left(\frac{11}{17}\epsilon^{-\frac{2(\pb+1)}{2\pb+13}}+\frac{2(\pb+1)}{17}\epsilon^\frac{11}{2\pb+13}\right)}

\newcommand{\energyeqdev}{E_{\rm{eq}}^{+\rm{dev}}=\Gamma\delta_D^{-\frac{7\pb+29}{2\pb+13}}E_{\rm{\rm{eq,N}}}^{+\rm{dev}}}

\newcommand{\energyeqN}{E_{\rm{\rm{eq,N}}}=\left(\xi^{11}\gamma_m^{11(2-\pb)}\frac{17^{2\pb+13}\pi^{\pb+1}\curlc^{3(\pb+1)}c^{4\pb+37}m_e^{\pb+12}F_p^{3\pb+14}d_L^{2(3\pb+14)}\eta^{5(\pb+1)}f_V^{2(\pb+1)}}{2^{7\pb-4}(\pb+1)^{2(\pb+1)}3^5 11^{11} \sqrt{3}q_e^{2(2\pb+13)}\nu_p^{2\pb+13}(1+z)^{5\pb+27}f_A^{3(\pb+1)}}\right)^\frac{1}{2\pb+13}\left(\frac{2\pb+13}{17}\right)}

\newcommand{\energyeq}{E_{\rm{eq}}=\Gamma\delta_D^{-\frac{7\pb+29}{2\pb+13}}E_{\rm{\rm{eq,N}}}}

\newcommand{\betaeqNtildedev}{
    \tilde{\beta}_{\rm{eq,N+dev}}
    \equiv \frac{\beta_{\rm{eq,N+dev}}}{(\Gamma - 1)^\frac{2 - \pb}{2\pb + 13}}
    = \frac{1+z}{ct} \left(\xi\epsilon \mu^{2-\pb} \chi_e^{2-\pb} \frac{(\pb+1)\curlc^{\pb+5} c F_p^{\pb+6} d_L^{2(\pb+6)} \eta^{\frac{5}{3}(\pb+5)}}{2^{\pb+2}11\sqrt{3}\pi^{\pb+7}m_e^{\pb+6}\nu_p^{2\pb+13}(1+z)^{3\pb+19}f_A^{\pb+5}f_V}\right)^\frac{1}{2\pb+13}
}

\newcommand{\betaeqNtilde}{
    \tilde{\beta}_{\rm{\rm{eq,N}}}
    \equiv \frac{\beta_{\rm{\rm{eq,N}}}}{(\Gamma - 1)^\frac{2 - \pb}{2\pb + 13}}
}

\newcommand{\betaeqNdev}{\beta_{\rm{eq,N+dev}} \equiv \frac{(1+z) R_{\rm{\rm{eq,N}}}^{+\rm{dev}}}{ct}
    = \frac{1+z}{ct} \left(\xi\epsilon \gamma_m^{2-\pb} \frac{(\pb+1)\curlc^{\pb+5} c F_p^{\pb+6} d_L^{2(\pb+6)} \eta^{\frac{5}{3}(\pb+5)}}{2^{\pb+2}11\sqrt{3}\pi^{\pb+7}m_e^{\pb+6}\nu_p^{2\pb+13}(1+z)^{3\pb+19}f_A^{\pb+5}f_V}\right)^\frac{1}{2\pb+13}
}

\newcommand{\betaeqN}{\beta_{\rm{\rm{eq,N}}} \equiv \frac{(1+z) R_{\rm{\rm{eq,N}}}}{ct}}

\newcommand{\thetactildedev}{\tilde{\theta}_c \equiv \tilde{\beta}_{\rm{eq,N+dev}}^{-\frac{2\pb + 13}{3(\pb + 6)}}}

\newcommand{\thetacdev}{\theta_c \equiv \beta_{\rm{eq,N+dev}}^{-\frac{2\pb + 13}{3(\pb + 6)}} = \tilde{\theta}_c (\Gamma - 1)^\frac{\pb - 2}{3(\pb + 6)}}

\newcommand{\thetactilde}{\tilde{\theta}_c \equiv \tilde{\beta}_{\rm{\rm{eq,N}}}^{-\frac{2\pb + 13}{3(\pb + 6)}}}

\newcommand{\thetac}{\theta_c \equiv \beta_{\rm{\rm{eq,N}}}^{-\frac{2\pb + 13}{3(\pb + 6)}} = \tilde{\theta}_c (\Gamma - 1)^\frac{\pb - 2}{3(\pb + 6)}}

\newcommand{\fourvelConstraint}{f(u) \equiv \left(\sqrt{1 + u^2} -u\cos\theta\right) \left(1 - \frac{1}{1+u^2}\right)^{-\frac{2\pb + 13}{6(\pb + 6)}} - \tilde{\theta}_c \left(\sqrt{1 + u^2} - 1\right)^\frac{\pb - 2}{3(\pb + 6)} = 0}

\newcommand{\onAxisLB}{u_{\text{on},\text{lb}} = (1 + \tilde{\theta}_c)^{-\frac{3(\pb + 6)}{2\pb + 13}}}

\newcommand{\offAxisUB}{u_{\text{off}, \text{ub}} = \left(\frac{\tilde{\theta}_c}{1-\cos\theta}\right)^\frac{3(\pb+ 6)}{2(\pb + 10)}}

\newcommand{\peakFreq}{\nu_p=\frac{\delta_D q_e B \gamma_e^2}{2\pi m_e c (1+z)}}

\newcommand{\peakSynchFlux}{F_p = \frac{(1+z) \delta_D^3 \sqrt{3} q_e^3 B N_e}{4\pi d_L^2 m_e c^2} \kappa^{1 - p}}

\newcommand{\peakBBFlux}{F_p = \frac{(1 + z)^2 \delta_D 2 m_e \gamma_e \nu_p^2 A}{\curlc d_L^2} \eta^{-\frac{5}{3}}}

\newcommand{\peakLorentz}{\gamma_e = \frac{\curlc F_p d_L^2 \eta^\frac{5}{3} \Gamma^2}{2\pi \nu_p^2 (1+z)^3 m_e f_A R^2 \delta_D}}

\newcommand{\magField}{B = \frac{8 \pi^3  m_e^3 c \nu_p^5 (1+z)^7 f_A^2 R^4 \delta_D}{\curlc^2 q_e F_p^2 d_L^4 \eta^\frac{10}{3} \Gamma^4}}

\newcommand{\numElec}{N_e = \frac{\curlc^2 c F_p^3 d_L^6 \eta^\frac{10}{3} \Gamma^4}{2 \sqrt{3} \pi^2 q_e^2 m_e^2 \nu_p^5 (1+z)^8 f_A^2 R^4 \delta_D^4} \kappa^{1-p}}

\newcommand{\numDens}{
    n_{\text{ext}} = \frac{3 \Gamma^2 N_e}{\pi R^3 f_\Omega}
} 

\newcommand{\electronEnergy}{
    E_e
    = N_e \Gamma \gamma_m m_e c^2
    = \left(\frac{\gamma_m}{\gamma_e}\right)^{2-\pb} \frac{N_e}{\kappa^{1-p}} \Gamma \gamma_e m_e c^2
    = \gamma_m^{2-\pb} \frac{\curlc^{\pb + 1} c^3 F_p^{\pb + 2} d_L^{2(\pb+2)} \eta^{\frac{5}{3}(\pb+1)} \Gamma^{2\pb+3}}{2^{\pb} \sqrt{3} \pi^{\pb+1} q_e^2 m_e^{\pb} \nu_p^{2\pb + 3} (1+z)^{3\pb+5} f_A^{\pb+1} R^{2(\pb + 1)} \delta_D^{\pb+3}}
}

\newcommand{\magFieldEnergy}{
    E_B
    = \Gamma^2 V \frac{B^2}{8\pi}
    = \frac{8 \pi^6 m_e^6 c^2 \nu_p^{10} (1+z)^{14} f_A^4 f_V \delta_D^2 R^{11}}{\curlc^4 q_e^2 F_p^4 d_L^8 \eta^\frac{20}{3} \Gamma^{10}}
}

\newcommand{\OoEderiv}{
    \frac{E_B}{\xi E_e} 
    = \frac{\epsilon_B}{\xi \epsilon_e} 
    = \frac{2(\pb + 1)}{11} \left(\frac{R}{R_{\rm{eq}}}\right)^{2\pb + 13}
}

\newcommand{\OoEradius}{R_{\rm{eq}}^{+\rm{dev}} \equiv R = R_{\rm{eq}} \epsilon^\frac{1}{2\pb + 13}}

\newcommand{\dynamics}{t=\frac{(1+z)R}{c\beta}(1-\beta\cos\theta)}

\newcommand{\constraintDeriv}{\frac{R}{R_{\rm{eq}}} \frac{R_{\rm{eq}}}{R_{\rm{\rm{eq,N}}}} = \frac{\beta}{\beta_{\rm{\rm{eq,N}}}} \Gamma \delta_D}

\newcommand{\constraintDerivNoteOne}{\frac{R}{R_{\rm{eq}}} = \epsilon^\frac{1}{2\pb + 13}}

\newcommand{\constraintDerivNoteTwo}{\frac{R_{\rm{eq}}}{R_{\rm{\rm{eq,N}}}} = \Gamma \delta_D^{-\frac{\pb + 5}{2\pb + 13}}}

\newcommand{\gammaBulkConstraintdev}{\delta_D = \epsilon^\frac{1}{3(\pb+6)} \left(\frac{\beta}{\beta_{\rm{\rm{eq,N}}}}\right)^{-\frac{2\pb + 13}{3(\pb + 6)}}
\equiv \left(\frac{\beta}{\beta_{\rm{\rm{eq,N}}}^{+\rm{dev}}}\right)^{-\frac{2\pb + 13}{3(\pb + 6)}}}

\newcommand{\gammaBulkConstraint}{\delta_D = \left(\frac{\beta}{\beta_{\rm{\rm{eq,N}}}}\right)^{-\frac{2\pb + 13}{3(\pb + 6)}}}

\newcommand{\fourvelconstraintbothfreq}{
    f(u)\equiv(\sqrt{1+u^2}-u\cos\theta)\left(1-\frac{1}{1+u^2}\right)^{-\frac{17}{48}}-\theta_c=0
}

\newcommand{\ReqNbothfreq}{R_{\rm{\rm{eq,N}}}^{+\rm{dev}}=
    \left[\xi \epsilon \left(\frac{\nu_m}{\nu_a}\right)^{\frac{(2-\pb)}{2}}
    \frac{3 \curlc^7 c F_p^8 d_L^{16} \eta^\frac{35}{3}}
    {2^4 11 \sqrt{3} \pi^9 m_e^8 \nu_p^{17}f (1+z)^{25} f_A^7 f_V}\right]^{\frac{1}{17}}
}

\newcommand{\Reqbothfreq}{
    R_{\rm{eq}}^{+\rm{dev}}=R_{\rm{\rm{eq,N}}}^{+\rm{dev}}\Gamma\delta_D^{-\frac{7}{17}}
}

\newcommand{\EeqNbothfreq}{E_{\rm{\rm{eq,N}}}^{+\rm{dev}}=
    \left[\xi^{11} \left(\frac{\nu_m}{\nu_a}\right)^{\frac{11(2-\pb)}{2}}
    \frac{{17}^{17} \pi^3 \curlc^9 c^{45} m_e^{14} F_p^{20} d_L^{40} \eta^{15} f_V^6}
    {2^{10} 3^{11} 11 \sqrt{3} q_e^{34} \nu_p^{17} (1+z)^{37} f_A^9}\right]^{\frac{1}{17}}
    \left[\frac{11}{17} \epsilon^{-\frac{6}{17}} + \frac{6}{17} \epsilon^{\frac{11}{17}}\right]
}

\newcommand{\Eeqbothfreq}{
    E_{\rm{eq}}^{+\rm{dev}}=E_{\rm{\rm{eq,N}}}^{+\rm{dev}}\Gamma\delta_D^{-\frac{43}{17}}
}

\newcommand{\electronEnergybothfreq}{E_e
    =N_e \Gamma \gamma_m m_e c^2
    =\left(\frac{\nu_m}{\nu_a}\right)^{\frac{2-\pb}{2}} \frac{N_e}{\kappa^{1-p}} \Gamma \gamma_e m_e c^2
    =\left(\frac{\nu_m}{\nu_a}\right)^{\frac{2-\pb}{2}}
    \frac{\curlc^3 c^3 F_p^4 d_L^8 \eta^{\frac{15}{3}} \Gamma^7}
    {2^2 \sqrt{3} \pi^3 q_e^2 m_e^2 \nu_p^6 (1+z)^{11} f_A^3 R^6 \delta_D^5}
}

\newcommand{\Etotbothfreq}{E_{tot}
    = \xi E_e + E_B
    =E_{\rm{eq}}\left[
    \frac{11}{17} \left(\frac{R}{R_{\rm{eq}}}\right)^{-6}
    + \frac{6}{17} \left(\frac{R}{R_{\rm{eq}}}\right)^{11} 
    \right]
}

\newcommand{\NewtonianGammaM}{
    \gamma_{m} = \frac{9}{32} \mu \chi_e \beta_{\rm sh}^2 = \frac12 \mu \chi_e \beta^2
}

\newcommand{\NewtonianGammaMsolved}{
    \gamma_{m} = 
    \left[ \xi \epsilon \mu^{\frac{2\pb+13}{2}} \chi_e^{\frac{2\pb+13}{2}} \frac{ (\pb+1)\curlc^{\pb+5} F_p^{\pb+6} d_L^{2(\pb+6)}\eta^{\frac{5}{3}(\pb+5)}}
    {2^{\frac{4\pb+17}{2}} 11\sqrt{3}\pi^{\pb+7}m_e^{\pb+6}\nu_p^{2\pb+13} (1+z)^{\pb+6} f_A^{\pb+5}f_V c^{2\pb + 12} t^{2\pb+13}} \right]^{\frac{2}{4\pb+9}}
}

\newcommand{\onAxisLBbothfreq}{u_{\text{on},\text{lb}} = (1 + \theta_c)^{-\frac{24}{17}}}

\newcommand{\offAxisUBbothfreq}{u_{\text{off}, \text{ub}} = \frac{\theta_c}{1-\cos\theta}}

\newcommand{\NewtonianGammaMsolvedbothfreq}{
    \gamma_{m} = 
    \left[ \xi \epsilon \mu^{\frac{17}{2}} \chi_e^{\frac{17}{2}} \left( \frac{\nu_m}{\nu_a} \right)^{\frac{2-\pb}{2}} \frac{3 \curlc^{7} F_p^{8} d_L^{16}\eta^{\frac{35}{3}}}
    {2^{\frac{25}{2}} 11\sqrt{3}\pi^{9}m_e^{8}\nu_p^{17} (1+z)^{8} f_A^{7} f_V c^{16} t^{17}} \right]^{\frac{2}{17}}
}

\begin{document}

\title{A Self-Consistent Framework for Synchrotron Equipartition Analysis}

\newcommand{\utah}{\affiliation{Department of Physics \& Astronomy, University of Utah, Salt Lake City, UT 84112, USA}}
\newcommand{\UA}{\affiliation{Steward Observatory, University of Arizona, 933 North Cherry Avenue, Tucson, AZ 85721-0065, USA}}

\author[orcid=0009-0007-3919-8439, sname='Rohde']{Coleman Rohde}
\utah
\email[show]{rohdecoleman@gmail.com}  

\author[orcid=0000-0003-1792-2338, sname="Laskar"]{Tanmoy Laskar}
\utah
\email{tanmoy.laskar@utah.edu}

\author[orcid=0000-0003-4537-3575, sname="Noah"]{Noah Franz}
\UA
\email{nfranz@arizona.edu}

\author[orcid=0009-0008-5392-4190, sname='Farley']{Gavin Farley}
\utah
\email{gavin.farley@utah.edu}

\author[orcid=0000-0003-0528-202X, sname="Christy"]{Collin Christy}
\UA
\email{collinchristy@arizona.edu}

\author[orcid=0000-0002-8297-2473,gname=Kate,sname=Alexander]{Kate~D.~Alexander}
\UA
\email{kdalexander@arizona.edu}

\begin{abstract}
Determining the energy, size, and velocity of synchrotron-emitting outflows is essential for testing models of their formation and evolution, but these quantities are often poorly constrained by observations alone. Equipartition analysis, therefore, provides a widely used framework for estimating these properties. Prior works have developed refinements to account for additional physical effects and other sources of energy (e.g., self-absorption, hot protons, and deviations from strict equipartition); however, these corrections are typically applied independently of one another, resulting in internal inconsistencies. In this work, we derive a self-consistent equipartition framework that accounts for the interdependence of various correction factors for Newtonian outflows and on- and off-axis relativistic jets. We implement our framework in an easy-to-use, publicly available code and apply it to study the tidal disruption events ASASSN-19bt and AT2019dsg, fast X-ray transient EP240414a, and active galactic nucleus J0231-0433. The interdependence of the corrections can increase energy estimates by a factor of $\sim$5, suggesting that the energies of other synchrotron sources may be similarly underestimated in the literature. These results indicate that simultaneously incorporating these correction factors is essential for determining accurate outflow properties and constraining launch mechanisms. 

\end{abstract}


\section{Introduction}
Astrophysical sources spanning many orders of magnitude in size and energy produce synchrotron radiation, from stellar-mass compact objects to radio galaxies and galaxy clusters. The resulting radio-to-X-ray spectra are commonly characterized by broken power laws that arise from a non-thermal population of relativistic electrons with a power-law Lorentz factor distribution \citep{gs65,bel78,bo78}. The characteristic synchrotron frequency ($\nu_m$), associated with the minimum electron Lorentz factor ($\gamma_m$), typically marks the peak of the optically thin emission. At sufficiently low frequencies, however, synchrotron self-absorption (SSA) renders the source optically thick below the absorption frequency $\nu_a$. In compact sources, the intrinsic synchrotron peak is often hidden beneath SSA such that the observed spectral peak corresponds to $\nu_a$, although in some systems both $\nu_m$ and $\nu_a$ can be directly identified in the spectral energy distribution (SED).

Equipartition \citep[e.g.,][]{bur56,pac70,mof75,Scott_Readhead_1977,bk05} provides a simple and robust framework for translating such synchrotron spectra into physical properties of the emitting region. By assuming that the energy stored in relativistic particles and magnetic fields is close to equipartition, one can estimate the source size, magnetic field strength, and minimum total energy with only a small number of observables. This technique has been applied across a remarkable range of astrophysical systems, including X-ray binaries \citep{fb19}, gamma-ray burst afterglows \citep{fwk00,Barniol_Duran_Nakar_Piran_2013}, supernova remnants \citep{che98,aua+12,upa18}, tidal disruption events \citep{zbs+11,abg+16}, star-forming galaxies \citep{con92}, radio galaxies and Active Galactic Nuclei (AGN) \citep{mil80,cb96}, and galaxy clusters \citep{gf04}. While the classical formulation assumes a stationary or mildly relativistic source, subsequent work has extended the formalism to relativistic outflows \citep{Barniol_Duran_Nakar_Piran_2013} and, more recently, to arbitrary viewing geometries \citep{Matsumoto_Piran_2023}.

However, a number of additional physical effects, which can significantly modify the inferred parameters, have typically been treated in a non-uniform and often ad hoc manner. For example, when $\nu_m < \nu_a$, the emission from electrons at $\gamma_m$ is self-absorbed and does not directly contribute to the observed spectrum, requiring additional assumptions about the electron distribution. In cases where both $\nu_m$ and $\nu_a$ are observed, one can instead express the system of equations in terms of the ratio of $\nu_m/\nu_a$, thereby removing the need to infer the minimum electron Lorentz factor from first principles. Similarly, the total energy budget may include contributions beyond the radiating electrons and magnetic field, such as energy carried by hot protons or thermal particle populations \citep[e.g.,][]{ Margalit_Quataert_2021,Margalit_Quataert_2024_nonArxiv}. Observations of a cooling break can further constrain the magnetic field strength and, consequently, the degree to which the system deviates from equipartition. While individual studies have incorporated one or more of these corrections, they have generally not been implemented within a consistent framework. Moreover, these effects are not independent; they are coupled through the underlying physical parameters of the system and therefore require a self-consistent treatment. For example, \cite{Cowie_Fender_2026} has explored how deviations from equipartition corrections are effected by both hot proton and $\nu_m$ electron corrections. 

In this work, we build upon existing generalized equipartition frameworks \citep[e.g.,][]{Barniol_Duran_Nakar_Piran_2013, Matsumoto_Piran_2023, Cowie_Fender_2026} and derive equipartition equations for both relativistic and non-relativistic outflows accounting for, simultaneously and self-consistently, the energy of SSA suppressed electrons, the energy of non-radiating components, and out of equipartition scenarios which give an estimate of the energy above the absolute minimum. We include contributions from non-radiating components of the outflow, such as energy in hot protons, without relying on assumptions of small electron energy fractions or strict equipartition. We consistently account for the role of electrons radiating at $\nu_m$, both in cases where this frequency is directly observed and when it is hidden by self-absorption. We extend the work of \cite{Cowie_Fender_2026} by fully constraining deviations from equipartition with the magnetic and (non-thermal) electron energy density fractions so that the relative amount of energy in hot protons is not a required parameter for equipartition analysis. We further relax common approximations adopted in relativistic treatments by solving for the bulk Lorentz factor numerically, without assuming $\Gamma \gg 1$ or small viewing angles. In the Newtonian regime ($\Gamma\beta\lesssim1$), we similarly solve for the minimum electron Lorentz factor and shock velocity self-consistently, rather than adopting fixed parameterizations.

We apply our method to two well-studied TDEs, ASASSN-19bt and AT2019dsg, a fast X-ray transient, EP240414a, and an AGN, J0231-0433, and demonstrate that our framework reproduces previous results where applicable when the corresponding assumptions are adopted. For the TDEs studied, we then explore the impact of including all corrections simultaneously, highlighting the differences in the inferred physical parameters. To facilitate broader adoption of our new framework, we provide a publicly available code\footnote{https://github.com/rohdog2003/equipartition} that implements the full set of equations and corrections in a uniform manner \citep[][]{rohde_2026_20835476}.  Our goal is to provide a flexible and robust tool for the community that enables consistent interpretation of radio observations across a wide range of astrophysical sources.

This paper is organized as follows. In Section~2, we present the derivation of the generalized equipartition framework. In Section~3, we compare our results with previous analyses and discuss the impact of the various corrections. In Section~4, we summarize our findings and outline their implications. The code developed in this work is publicly available on GitHub, and we hope it will serve as a useful resource for future studies of synchrotron-emitting transients and outflows.

\section{A comprehensive equipartition framework} 
\subsection{Setup and basic assumptions}
From some progenitor at redshift $z$ and luminosity distance $d_L$, a fluid shock of radius $R$ propagates into an environment with density, $n_{\rm ext}$. Electrons swept up and accelerated by the shock produce synchrotron emission in the presence of the post-shock magnetic field ($B$), with an SED peaking at frequency $\nu_p$ with peak flux, $F_p$. By considering synchrotron and black body theory, an expression for the energy of the system in terms of the radius $R$ can be found and minimized to obtain a lower limit on the energy. All that must be inferred is the observer angle $\theta$ and geometry of the outflow, the latter parametrized by the area-filling factor $f_A=A/(\pi R^2/\Gamma^2)$, volume-filling factor $f_V=V/(\pi R^3/\Gamma^4)$, and solid-angle-filling factor $f_\Omega=\Omega/(\pi/\Gamma^2)$, where $\Omega$, $V$, and $A$ are the  solid angle, volume, and projected area of the emitting region, respectively. The bulk Lorentz factor of the post-shock fluid, $\Gamma=1/\sqrt{1-\beta^2}$ and Doppler factor\footnote{\cite{Barniol_Duran_Nakar_Piran_2013} take $\delta_D\approx1.4\Gamma$ (see \citet{Matsumoto_Piran_2023}).} $\delta_D=[\Gamma(1-\beta\cos\theta)]^{-1}$ are constrained by assuming a relation between $R$ and observer-frame time, $t$. 

We assume a fraction $\epsilon_e$ of the post-shock fluid energy is provided to electrons accelerated into a power law injection distribution, $N_{\text{inj}}(\gamma)\propto H(\gamma-\gamma_m)\gamma^{-p}$, where $H$ is the Heaviside step function, 
\begin{equation}
    \gammaM
    \label{gammaMequationlabel}
\end{equation}
is the minimum Lorentz factor (see Section~\ref{sec:newtoniangammam} for a discussion of $\gamma_m$ in the Newtonian regime), $\mu$ is the mean composition of the ambient medium, and we have defined $\chie$. If equation (\ref{gammaMequationlabel}) gives $\gamma_m<2$, we set $\gamma_m=2$. The spectrum peaks at the greater of the critical frequencies $\nu_m$ (the peak synchrotron frequency of electrons radiating with Lorentz factor $\gamma_m$) and $\nu_a$ (the frequency below which synchrotron radiation is effectively self absorbed). 

We constrain $B$, the number of swept-up electrons ($N_e$), and the Lorentz factor of the electrons radiating primarily at $\nu_p$ by matching the synchrotron and Rayleigh-Jeans peak fluxes. The Lorentz factor ($\gamma_e$) of electrons radiating at the peak is related to the observed peak of the synchrotron spectrum ($\nu_p$) via 

\begin{equation}
    \peakFreq, 
    \label{peakFreqLabel}
\end{equation} while the synchrotron flux at $\max{(\nu_a,\nu_m)}$ is given by
\begin{equation}
    \peakSynchFlux,
    \label{peakSynchFluxLabel}
\end{equation}
where
\begin{equation}
    \kappaeq \qquad
    \label{kappaeqlabel}
\end{equation}
\noindent corrects\footnote{For $\nu_m<\nu_a$ (corresponding to $\kappa > 1$), electrons which radiate at $\nu_m$ significantly contribute to the total non-thermal electron energy, even though their observed emission is suppressed by SSA.} for the effects of self-absorption when $\nu_m < \nu_a$. 
\noindent Like \cite{Barniol_Duran_Nakar_Piran_2013} and \cite{Matsumoto_Piran_2023} we neglect the $p$-dependence\footnote{See \cite{Wijers_Galama_1999} for a more precise treatment.} of the numerical factors of equations (\ref{peakFreqLabel}) and (\ref{peakSynchFluxLabel}) and use approximate values for $p>2$. Following \cite{Matsumoto_Piran_2023}, the peak flux from a black-body of emitting area $A$ in the Rayleigh-Jeans regime, extended to match the synchrotron spectrum, is given by
\begin{equation}
    \peakBBFlux,
    \label{peakBBFluxLabel}
\end{equation}
where
\begin{equation}
    \etaeq
\end{equation}
The numerical factor,  
\begin{equation}
    \shenzhangfactor
    \label{shenzhangfactorlabel}
\end{equation} 
where $G$ is the gamma function, is required to precisely match\footnote{\cite{Barniol_Duran_Nakar_Piran_2013} and \cite{Matsumoto_Piran_2023} take $\curlc\sim 3$. $\curlc$ is discontinuous across $\nu_m=\nu_a$, causing a slight discontinuity in the estimated quantities from equipartition analysis when both $\nu_a$ and $\nu_m$ are inferred from the observed spectral energy distribution and seen to cross each other (see also Section~\ref{bothfreqsection}).} the Rayleigh-Jeans and synchrotron peak fluxes in both cases of $\nu_a<\nu_m$ and $\nu_m<\nu_a$ \citep{Shen_Zhang_2009}. 

Equations (\ref{peakFreqLabel}), (\ref{peakSynchFluxLabel}), and (\ref{peakBBFluxLabel}) can now be solved to obtain expressions for $\gamma_e$, $N_e$, and $B$ as
\begin{equation}
    \peakLorentz
\end{equation}
\begin{equation}
    \magField
\end{equation}
\begin{equation}
    \numElec
    \label{numElecLabel}
\end{equation}
The factor $\kappa^{1-p}$ in equation (\ref{numElecLabel}) is also discussed in \cite{Cendes_Alexander_Berger_Eftekhari_Williams_Chornock_2021} and accounts for\footnote{\cite{Cendes_Alexander_Berger_Eftekhari_Williams_Chornock_2021} identify the factor of $\kappa^{1-p}$, which can be rewritten as $(\gamma_e/\gamma_m)^{1-p}$ for $\nu_a<\nu_m$, as a Newtonian correction; however, it is actually a correction due to electrons radiating at $\nu_m$.} electrons radiating at $\nu_m$. The electron energy ($E_e$) and magnetic field energy ($E_B$) in the rest frame are then given by\footnote{Throughout this work, the electron energy always refers to energy in relativistic (non-thermal) electrons.}
\begin{equation}
    \electronEnergy
    \label{electronEnergyLabel}
\end{equation} 
\begin{equation}
    \magFieldEnergy,
    \label{magFieldEnergyEqLabel}
\end{equation}
where the definition
\begin{equation}
    \pbar
\end{equation}
allows us to concisely write the electron energy expression in equation (\ref{electronEnergyLabel}) for both orderings of $\{\nu_m,\nu_a\}$, since the factor $(\gamma_m/\gamma_e)^{2-\pb}$ that accounts for\footnote{This factor originates from the $\kappa^{1-p}$ factor in $N_e$, which we divide out in equation (\ref{electronEnergyLabel}) to facilitate comparisons with \cite{Barniol_Duran_Nakar_Piran_2013}} the energy of electrons radiating at $\nu_m$  \citep{Barniol_Duran_Nakar_Piran_2013} disappears when $\nu_a<\nu_m$.
Finally the number  density of the ambient medium can be found by dividing the number of electrons in the shock by the volume the shock has swept \citep[][]{Matsumoto_Piran_2023} as
\begin{equation}
    \numDens.
\end{equation}
This expression is different from the expressions of \cite{Christy_et_al._2024} and \cite{Cendes_Alexander_Berger_Eftekhari_Williams_Chornock_2021}, who instead estimate the number density of the emitting region as $n_\text{int}=N_e/V$ and then relate $n_\text{int}$ to the number density of the ambient medium via shock compression, $n_\text{ext}=n_\text{int}/(4\Gamma^2)$. 

\subsection{Hot Proton Corrections}
Next, we account for energy elsewhere in the system, i.e., energy that is not in the magnetic field or relativistic (non-thermal) electrons. We assume all of the components of the total energy $E$ occupy the same volume $V$ of the emitting region so that the magnetic energy fraction, $\epsilon_B=E_B/E$, electron energy fraction $\epsilon_e=E_e/E$, and the fraction $\epsilon_o=E_o/E$ of energy elsewhere $E_o$ add to unity $\epsilon_e+\epsilon_{o}+\epsilon_B=1$. This energy elsewhere, which can be rewritten in terms of the electron energy $E_o=(1-\epsilon_e-\epsilon_B)\epsilon_e^{-1} E_{\text{e}}$, can contribute significantly to the total energy. We call the corrections, which account for this energy elsewhere (including, e.g., hot protons and thermal electrons), ``hot proton corrections'' in analogy with \cite{Barniol_Duran_Nakar_Piran_2013}, who assumed that almost all of the energy elsewhere in the system was contained in hot protons; however, our parameterization allows us to extend that definition to all other particles receiving a share of the post-shock energy. In analogy with \cite{Barniol_Duran_Nakar_Piran_2013}, we define
\begin{equation}
    \xiHP > 1
\end{equation}
to parameterize the importance of energy elsewhere in the system, relative to energy in non-thermal electrons and magnetic fields. 
When all the energy of the shock is in the electrons and the magnetic field, i.e., $\epsilon_e+\epsilon_B=1$, $\xi=1$ and this correction factor drops out of the equations. Unlike the approximation for the hot proton parameter $\xi=1+\epsilon_e^{-1}$ in \cite{Barniol_Duran_Nakar_Piran_2013}, which only holds for $\epsilon_e\ll1$ and $\epsilon_B\ll\epsilon_e$, our expression for $\xi$ holds for any value of $\epsilon_e$ and $\epsilon_B$. 

\subsection{Equipartition Radius and Energy}
To estimate the equipartition radius and equipartition energy\footnote{Following common terminology, we use the term ``equipartition energy'' to refer to the minimum energy, and the term ``deviation from equipartition'' to refer to systems where the energy is greater than this minimum energy.}, we combine the relativistic electron energy, magnetic field energy, and energy anywhere else in the system $E_o$. Substituting in the equations (\ref{electronEnergyLabel}) and (\ref{magFieldEnergyEqLabel}) and minimizing the resulting expression relative to the radius yields the total energy in terms of a characteristic equipartition radius $R_{\rm eq}$,
    \Etot
where the minimizer (or ``equipartition'') radius is
\begin{equation}
    \Req,
    \label{Reqeqn}
\end{equation}
\begin{equation}
    \ReqN,
\end{equation}
and the minimum (or ``equipartition'') energy is, 
\begin{equation}
    \energyeq,
\end{equation}
\begin{equation}
    \energyeqN
\end{equation}

\subsection{Solutions for the Lorentz factor}
\label{sec:Lorentzfactorlabel}
\begin{figure}
    \centering
    \includegraphics[width=0.8\linewidth]{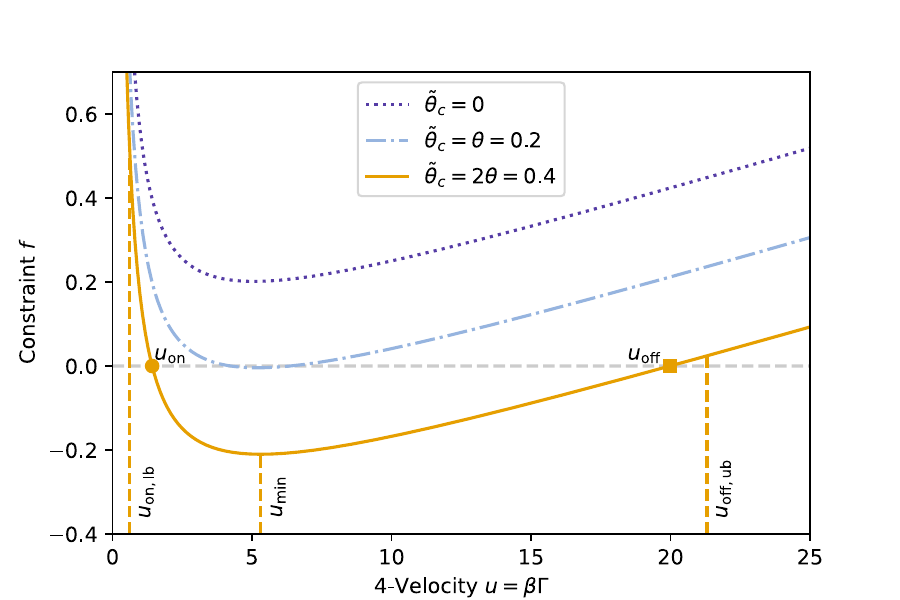}
    \caption{The on axis solution (circle), bounded below by $u_{\text{on,lb}}$ and above by $u_{\text{min}}$, and the off-axis solution (square), bounded below by $u_{\text{min}}$ and above by $u_{\text{off,ub}}$, exist for (approximate) critical angles $\tilde{\theta}_c\approx\theta_c$ larger than the observer angle $\theta$ i.e., for $\tilde{\theta}_c\gtrsim \theta$. $\theta=0.2$ is fixed here as an example.}
    \label{plotfFigure}
\end{figure}

We now consider the impact of the viewing geometry on the equipartition quantities. The range of viewing angles that allow relativistic solutions is bounded above by a critical maximum viewing angle given by \cite{Matsumoto_Piran_2023}
\begin{equation}
    \thetac,
\end{equation}
where the apparent shock velocity\footnote{In the Newtonian case, $\beta_{\rm eq,N}$ corresponds to the true shock velocity.},
\begin{equation}
    \betaeqN.
\end{equation}
Here we have defined
\begin{equation}
    \thetactilde
\end{equation}
and 
\begin{equation}
    \betaeqNtilde
\end{equation}
for convenience by extracting the Lorentz-factor dependence in $\beta_{\rm eq,N}$, which allows us to solve for $\Gamma$ self-consistently. Following \cite{Matsumoto_Piran_2023}, we relate the shock velocity fraction and time in the observer-frame as 
\begin{equation}
    \dynamics
    \label{dynamicsLabel}
\end{equation}
to constrain the Lorentz factor. It can be shown that equation (\ref{dynamicsLabel}) is equivalent to 
\begin{equation}
    \constraintDeriv
    \label{constraintDerivLabel}
\end{equation}
Using equation \ref{Reqeqn} and in equipartition ($R=R_{\rm eq}$), this yields the constraint 
\begin{equation}
    \gammaBulkConstraint.
    \label{gammaBulkConstraintLabel}
\end{equation}
These dynamical considerations now constrain the 4-velocity $u=\beta\Gamma$ as zeros of the function 
\begin{equation}
    \fourvelConstraint,
    \label{fourvelConstraint}
\end{equation}
from which $\Gamma=\sqrt{1+u^2}$ can be found . The function $f$ has a vertical asymptote at zero, a very clear global minimum at $u_{\text{min}}$, and two roots in the intervals $(0,u_{\text{min}})$ and $(u_{\text{min}},\infty)$ corresponding to the on- and off-axis solutions, respectively (provided $\tilde{\theta}_c\gtrsim\theta$; Figure~\ref{plotfFigure}). The on-axis solution is bounded below by 
\begin{equation}
    \onAxisLB, 
\end{equation}
where $u_{\text{on,lb}}>0$, and the off-axis solution is bounded above by 
\begin{equation}
    \offAxisUB.
\end{equation}
The 4-velocities, $u_{\text{on,lb}}$ and $u_{\text{off,ub}}$ can be determined by using algebraically tractable functions that are less than $f$ for $u\in(0,\infty)$ in the limits $\theta\ll 1$ and $\theta\gg 1$ respectively. After numerically finding $u_{\text{min}}$, these bounds can be used to reliably and quickly find the roots of $f$ numerically via Brent's method \citep[][]{brent_1972} in the intervals $(u_{\text{on,lb}}, u_{\text{min}})$ and $(u_{\text{min}}, u_{\text{off,ub}})$. This numerical approach does not require the assumptions that $\Gamma\gg 1$ and $\theta\ll 1$ of the analytical approach in \cite{Matsumoto_Piran_2023}.

\subsection{Deviations from Equipartition}\label{deviationssection}
Having discussed equipartition quantities, we now allow the system to deviate from equipartition (e.g., in cases where an independent estimate for $\epsilon_B$ is available). Working from equation (16) our above framework easily allows for this extension by the inclusion of a dimensionless out-of-equipartition parameter, which we define next. Writing $E_{\rm tot}=\xi E_e + E_B$ as in equation (16), the relative ratio of the two energy terms 
\begin{equation}
    \OoEderiv
\end{equation}
deviates from about unity (more precisely, $2(\pb+1)/11$) when the system is out of equipartition. Since equipartition analysis is simply an energy minimization argument, an ``in equipartition radius'' is just a radius which minimizes the total energy and so $E_o$ must be accounted for (through $\xi$) in this minimization. The corresponding radius 
\begin{equation}
    \OoEradius
    \label{OoEradiusLabel}
\end{equation}
is now smaller (for $\epsilon_B\ll1$), and the parameter 
\begin{equation}
    \epsOoE = \left(\frac{R}{R_{\rm eq}}\right)^{2\pb+13}
\end{equation}
quantifies the deviation of the system from equipartition. With the substitution $\xi=1+\eta\equiv1+E_o/E_e$ we recover Equation (C8) of \cite{Cowie_Fender_2026}. Hot protons and electrons that radiate at $\nu_m$ both affect the balance of energy between the electrons and the magnetic field and therefore modify deviations from equipartition (DFE). For this reason, our DFE parameter differs from the expression $\epsilon=(11/6)(\epsilon_B/\epsilon_e)$ (e.g., \citealt{Christy_et_al._2024}, \citealt{Barniol_Duran_Nakar_Piran_2013}) as it now considers all of these corrections self-consistently. Equipartition is achieved for $\epsilon=1$ where the magnetic energy density fraction is fully constrained $\epsilon_B=\frac{2(\pb+1)}{2\pb+13}$, with the magnetic energy density comprising roughly a third of the total post-shock energy density. 
When hot protons become unimportant $\xi=1$ we recover the usual approximately equal energy between the electrons and magnetic field of in-equipartition calculations. 

The corresponding expressions for the equipartition radius are 
\begin{equation}
    \Reqdev
    \label{ReqLabel}
\end{equation}
and
\begin{equation}
    \ReqNdev.
    \label{ReqNLabel}
\end{equation}
The total energy can be found by substituting equation (\ref{OoEradiusLabel}) into equation (17), yielding
\begin{equation}
    \energyeqdev
    \label{energyeqLabel}
\end{equation}
and
\begin{equation}
    \footnotesize\energyeqNdev
    \label{energyeqNLabel}
\end{equation}

The apparent velocity is now given by
\begin{equation}
    \betaeqNdev
    \label{betaEqNdevLabel}
\end{equation}
where we have defined
\begin{equation}
    \betaeqNtildedev. 
    \label{betaeqNtildeLabel}
\end{equation}
When the system is not in equipartition, the critical angle becomes 
\begin{equation}
    \thetacdev
\end{equation}
where
\begin{equation}
    \thetactildedev
    \label{thetactildedevlabel}
\end{equation}
For DFE calculations equation (\ref{thetactildedevlabel}) must be substituted into equation (\ref{fourvelConstraint}).
The Doppler parameter can be written (by substituting $\constraintDerivNoteOne$ and $\constraintDerivNoteTwo$ into equation \ref{constraintDerivLabel}) as
\begin{equation}
    \gammaBulkConstraintdev. 
    \label{gammaBulkConstraintdevLabel}
\end{equation}
In summary, all out-of-equipartition effects in the calculation of the Lorentz factor (or Doppler parameter) can be subsumed into corrections to the apparent velocity. Finally, equation (\ref{gammaBulkConstraintdevLabel}) rewritten in terms of the 4-velocity yields equation (\ref{fourvelConstraint}) in the same form as before, but with $\delta_D$ defined as in equation \ref{gammaBulkConstraintdevLabel}.

Equations \ref{ReqNLabel}, \ref{energyeqNLabel}, and \ref{gammaBulkConstraintdevLabel} now include hot proton, electrons radiating at $\nu_m$, and DFE corrections simultaneously and self-consistently. For $\epsilon=\xi=1$, $\pb=2$, and $\curlc=3$, these equations reduce to the expressions of \cite{Matsumoto_Piran_2023}. Due to the effect of the hot proton corrections on the DFE corrections, $\epsilon\propto \epsilon_B/(1-\epsilon_B)$ instead of $\epsilon\propto \epsilon_B/\epsilon_e$ as in \cite{Barniol_Duran_Nakar_Piran_2013}. Therefore, compared to previous calculations, $\epsilon$ is smaller in our work by an order of magnitude or more for typical values of the electron energy density fraction $\epsilon_e\lessapprox0.1$ and magnetic energy density fraction $\epsilon_B\ll1$. This raises the equipartition energy considerably for small values of $\epsilon_B$ where the $\epsilon^{-\frac{2(\pb+1)}{2\pb+13}}$ term dominates. Though a smaller effect, for large values of $p$, in the case $\nu_m < \nu_a$, the term becomes even larger as the exponent grows in magnitude. Interestingly, for out of equipartition calculations the consideration of hot protons has no effect on the equipartition radius, which depends only on the product $\xi\epsilon=\frac{11}{2(\pb+1)}\frac{\epsilon_B}{\epsilon_e}$ which is independent of $\xi$. Similarly, for $\epsilon_B\approx1$ and $\epsilon_e\ll1$ (i.e. $\epsilon\gg1$) the consideration of hot protons has no effect on the equipartition energy. Our expressions for the equipartition radius and energy are very similar to the corresponding equations (D18) and (D19) of \cite{Sfaradi_et_al_2025}, except here we provide the precise $\pb$-dependent numerical prefactors, and, additionally, our definition of the DFE parameter $\epsilon$ includes the effect of hot protons.

\section{Minimum Lorentz Factor in the Newtonian Regime}
\label{sec:newtoniangammam}
In the Newtonian regime ($\Gamma=1$) it is possible to obtain an analytical expression for $\gamma_m$ using the Newtonian expression

\begin{equation}
    \NewtonianGammaM
    \label{NewtonianGammaM}
\end{equation}

\noindent where $\beta=\frac34\beta_{\rm sh}$ and $\beta_{\rm sh}$ are the lab-frame speeds of the post-shock fluid and the shock, respectively, and $\mu$ is the mean composition of the external medium. The factor of $\frac{9}{32}$ comes from the shock jump conditions. In the Newtonian regime, the true and apparent speeds of the post-shock fluid are the same, $\beta=\beta_{\rm{\rm{eq,N}}}^{+\rm{dev}}$. Thus, we can solve equations (\ref{betaEqNdevLabel}) and (\ref{NewtonianGammaM}) for the minimum Lorentz factor yielding

\begin{equation}
    \NewtonianGammaMsolved .
    \label{NewtonianGammaMsolvedlabel}
\end{equation}
For typical Newtonian outflows this estimate of $\gamma_m$ is usually smaller than $2$, in which case we set $\gamma_m=2$.

\section{Identification of both $\nu_a$ and $\nu_m$}\label{bothfreqsection}
In the case that $\nu_a$ and $\nu_m$ can both be identified in the spectra it is no longer necessary to estimate a $\gamma_m$ theoretically. The corrections from electrons radiating at $\nu_m$ can now instead be parameterized by the observed ratio\footnote{This is especially important when moving from a $\nu_m$ peak to a $\nu_a$ peak since if both observed critical frequencies are used in the equipartition analysis during the $\nu_m$ peak but a $\gamma_m$ is instead estimated during a $\nu_a$ peak then the estimated $\gamma_m$ can be inconsistent with the observed $\nu_m$ causing a large discontinuity in equipartition estimates where $\nu_a$ crosses $\nu_m$.} of the critical frequencies $\nu_m/\nu_a$. First we consider the electron energy 

\begin{equation}
    \electronEnergybothfreq.
\end{equation}

\noindent The exponents of many of the observed quantities (e.g., peak flux, peak frequency, and radius) are no longer $\pb$ dependent since the observed ratio $(\nu_m/\nu_a)^{\frac{2-\pb}{2}}$ is used in place of the estimated $(\gamma_m/\gamma_e)^{2-\pb}$. The magnetic field energy is unchanged in this case from equation (\ref{magFieldEnergyEqLabel}). 

The deviations from equipartiton parameter in this case $\epsilon=11\epsilon_B/(6\xi\epsilon_e)$ is no longer dependent on $\pb$ since the ratio of the energies, $E_B/(\xi E_e)$, as derived from the total energy,
\begin{equation}
    \Etotbothfreq
\end{equation}

\noindent is no longer $\pb$ dependent, taking on the usual form as in \cite{Matsumoto_Piran_2023}. The equipartition radius and energy, including deviations from equipartition, are then found as
\begin{equation}
    \ReqNbothfreq,     \label{ReqNbothfreqLabel}
\end{equation}

\begin{equation}
    \Reqbothfreq,     \label{Reqbothfreqlabel}
\end{equation}

\begin{equation}
    \EeqNbothfreq,     \label{EeqNbothfreqLabel}
\end{equation}
and
\begin{equation}
    \Eeqbothfreq.    \label{EeqbothfreqLabel}
\end{equation}
Equations (\ref{ReqNLabel}), (\ref{ReqLabel}), (\ref{energyeqNLabel}), (\ref{energyeqLabel}) are equivalent to equations (\ref{ReqNbothfreqLabel}), (\ref{Reqbothfreqlabel}), (\ref{EeqNbothfreqLabel}), (\ref{EeqbothfreqLabel}) for a $\nu_m$ peaked spectrum where $\pb=2$.

The Lorentz factor is affected in this case. The equation constraining the Lorentz factor 
\begin{equation}
    \fourvelconstraintbothfreq
\end{equation}is found in much the same way as in Section \ref{sec:Lorentzfactorlabel}. A lower bound for the on axis solution
\begin{equation}
    \onAxisLBbothfreq
\end{equation}
and an upper bound for the off-axis solution 
\begin{equation}
     \offAxisUBbothfreq
\end{equation}
in this case once again allow the utilization of the Brent's method \citep[][]{brent_1972} for reliable root finding.

Finally, a solution of $\gamma_m$ for a Newtonian outflow
\begin{equation}
    \NewtonianGammaMsolvedbothfreq
\end{equation}
can also be found in this case.

\section{Comparisons to Previous Work}
We now compare our formalism to previously published equipartition analyses for two TDEs with well-sampled radio spectral energy distributions. We also perform an equipartition analysis of a Fast X-ray transient and an AGN. 

\subsection{ASASSN-19bt}\label{ASASSN-19btSubsectionLabel}
\begin{figure}[b]
    \centering
    \includegraphics[width=0.9\linewidth]{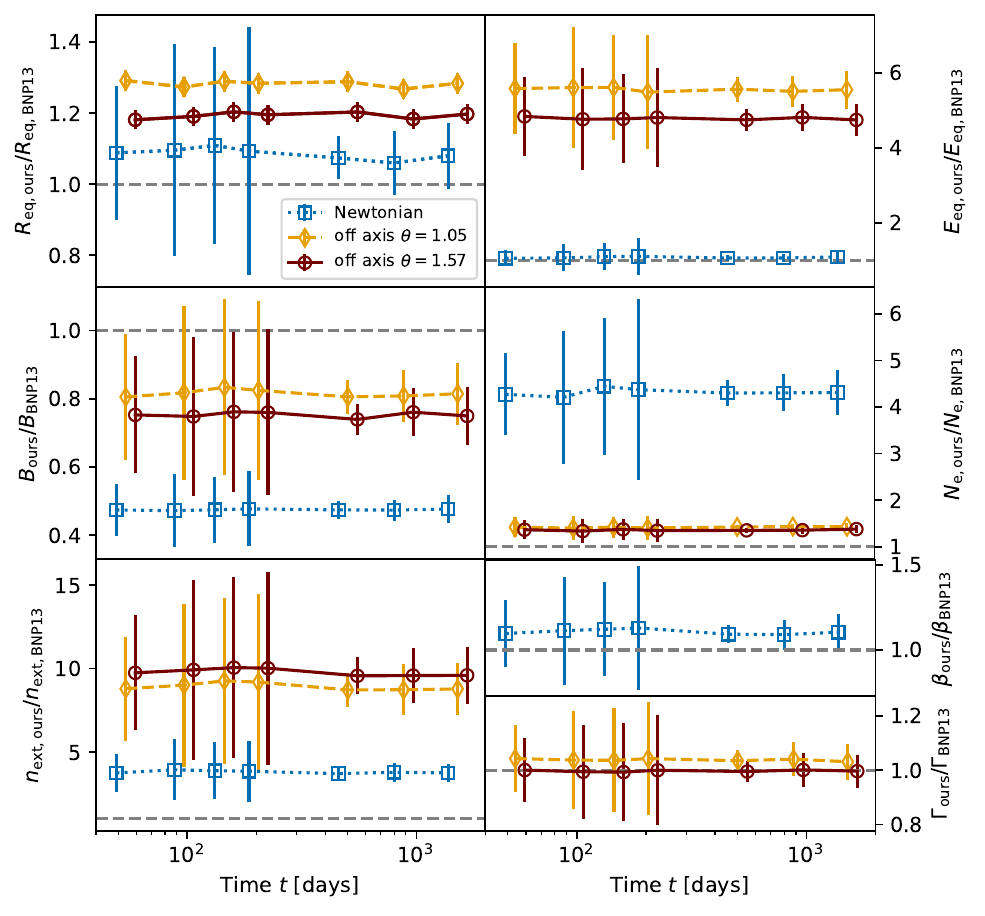}
    \caption{The ratio between equipartition quantities for the radio SEDs of the TDE ASASSN~19bt derived from our analysis and those reported by \cite{Christy_et_al._2024} (which used the \citet[][BNP13]{Barniol_Duran_Nakar_Piran_2013} formalism) for a Newtonian outflow model and two relativistic off-axis jet models. Here we assume the \cite{Planck_2018} Table 2 (TT, TE, EE + lowE + lensing + BAO) cosmology $H_0=67.66\text{km/s/Mpc}$, $\Omega_{m}=0.3111$, $\Omega_{\Lambda}=0.6889$, and the redshift $z=0.0262$, same microphysical, and geometric parameters in each case as in Table 2 of \cite{Christy_et_al._2024}. In the off-axis relativistic cases, $f_A=f_V=f_\Omega=1$. $\epsilon_e=0.1$ and $\epsilon_B=0.003$ for $\theta=1.05$. $\epsilon_e=0.1$ and $\epsilon_B=0.005$ for $\theta=1.57$. In the Newtonian case $\epsilon_e=0.1$, $\epsilon_B=0.2$, $f_A=1$, $f_V=0.36$ and $f_\Omega=4$.}
    \label{Christycomparisonfigure}
\end{figure}

Using our formalism and code, we perform an equipartition analysis of ASASSN-19bt in Table \ref{ChristyReproducedTable} and compare our analysis against that of \cite{Christy_et_al._2024} in Figure~\Ref{Christycomparisonfigure}. Using the expressions of \cite{Barniol_Duran_Nakar_Piran_2013}, \cite{Christy_et_al._2024} account for electrons radiating at $\nu_m$, hot protons, and deviations from equipartition by multiplying the individual factors for the corrections together. We use the $F_p$ and $\nu_p$ results of \cite{Christy_et_al._2024} which assumes $\nu_m<\nu_a$. Our equipartition model is similar to that of \cite{Christy_et_al._2024}, barring some important differences. First, we additionally consider the effects that electrons radiating at $\nu_m$ and hot protons have on DFE. This consideration, for small $\epsilon_B$ and typical $\epsilon_e\sim 0.1$, decreases the equipartition radius by a factor $0.1^{\frac{1}{2\pb+13}}\approx 0.88$ and increases the equipartition energy by a factor of $0.1^{-\frac{2(\pb+1)}{2\pb+13}}\approx 2.56$. Secondly, the analysis in \cite{Christy_et_al._2024} uses $\curlc\sim 3$ following \cite{Barniol_Duran_Nakar_Piran_2013}, while we consider the full expression for $\curlc\approx4.784$ for $p=2.8$, which decreases our estimate of $B$ by a factor of $9/\curlc^2\approx 0.39$, increases $N_e$ and $n_{\text{ext}}$ by a factor of $\curlc^2/9\approx 2.54$, increases the equipartition radius and apparent velocity by a factor of $(\curlc/3)^\frac{\pb+5}{2\pb+13}\approx 1.22$, and increases the equipartition energy by a factor of $(\curlc/3)^\frac{3(\pb+1)}{2\pb+13}\approx 1.33$. Following \cite{Matsumoto_Piran_2023}, we calculate the volume $V=R^3\Omega/3$ that the CNM occupies by considering the solid angle $\Omega$ subtended by the outflow and exclude the shock jump conditions factor of $4^{-1}$ such that our expression for $N_e$ is increased by a factor of $12f_V/f_\Omega\approx 3.82$ for the Newtonian case and $12f_V/f_\Omega\approx 9.36$ for the off-axis relativistic cases. Like \cite{Christy_et_al._2024} we also consider the effect of electrons radiating at $\nu_m$ on $N_e$. Following \cite{Matsumoto_Piran_2023} we exclude the factors of 4 in the Newtonian case introduced by \cite{Barniol_Duran_Nakar_Piran_2013} which arise from the spherical geometry in the Newtonian case. For the Newtonian case we set $\gamma_m=2$, as in \cite{Christy_et_al._2024}, since equation \ref{NewtonianGammaM} yields $\gamma_m<2$. For the off-axis relativistic models we also solve for the Lorentz factor numerically as done in \cite{Christy_et_al._2024}, but we additionally consider the effect of hot protons on DFE and how this affects the Lorentz factor. Although we consider a different definition for $\xi$ than \cite{Christy_et_al._2024}, for $\epsilon_e\ll1$ and $\epsilon_B\ll1$, the two definitions are very comparable in magnitude. We propagate the errors by assuming that the uncertanties on $F_p$ and $\nu_p$ are normally distributed with no covariance.

These combined effects  result in equipartition radii larger by a factor of $1.24$ on average, a $8.30\sigma$ systematic effect relative to the statistical uncertainty on this parameter (Figure~\ref{Christycomparisonfigure}). Our estimate of the off-axis relativistic energies is on average $5.17$ times greater at $6.54\sigma$. This difference is mostly accounted for by the effects of $\curlc$ and the effects of hot protons on DFE yielding a factor $0.1^{\frac{1}{2\pb+13}}(\curlc/3)^\frac{\pb+5}{2\pb+13}\approx 1.07$ increase for the equipartition radius and a factor $0.1^{-\frac{2(\pb+1)}{2\pb+13}}(\curlc/3)^\frac{3(\pb+1)}{2\pb+13}\approx 3.41$ increase in the equipartition energy. In the Newtonian case our equipartition radius and energy are not significantly different. This is partly due to our exclusion of factors of 4 which reduce our estimates in the Newtonian case. For the Newtonian case our magnetic field estimate is significantly smaller by a factor $2.13$ and our $N_e$ is larger by a factor $4.31$, both at $>5\sigma$. The decrease in our magnetic field estimates is due to $\curlc$ and the increased radius, since $B\propto R^{4}$, yielding a factor $(9/\curlc^2)^{-1}(R_{\text{ours}}/R_{\text{C24}})^{-4}\approx 1.79$ smaller in the Newtonian case. Similarly, since $N_e\propto R^{-4}$, the increase in our estimates of $N_e$ are partly explained by the factor, accounting for our exclusion of factors of 4, $4(9/\curlc^2)^{-1}(R_{\text{ours}}/R_{\text{C24}})^{-4}\approx 7.21$ in Newtonian case. We find a larger value for $n_{\text{ext}}$ by a factor of $9.35$ at $3.62\sigma$ in the off-axis relativistic case and $3.81$ at $3.59\sigma$ in the Newtonian case. This is explained by our exclusion of the shock jump condition factor and our different estimates of the radius and $N_e$, since $n_{\text{ext}}\propto N_e R^{-3}$, yielding an increase of $(12f_V/f_{\Omega})(N_{\text{e,ours}}/N_{\text{e,C24}})(R_{\text{ours}}/R_{\text{C24}})^{-3}\approx 8.75$ for the off-axis relativistic case and $(12f_V/f_{\Omega})(N_{\text{e,ours}}/N_{\text{e,C24}})(R_{\text{ours}}/R_{\text{C24}})^{-3}\approx 3.60$ in the Newtonian case. We find a nearly equivalent post-shock fluid speed and Lorentz factor for the Newtonian and off-axis relativistic cases, respectively. A summary of the significance of our results and the primary reasons for the differences are found in Table~\ref{significancetable}. Discrepancies between the expected factor and the actual factor difference observed between the analysis are due to minor differences not explicitly considered. For example, for $N_e$, in our expected factor, we do not account for the fact that $\gamma_e$ and $\gamma_m$ change between the two analyses so that actually the $\kappa=\max\{\gamma_e/\gamma_m, 1\}$ are different which contributes to the differences in $N_e$ as $N_e\propto\kappa^{1-p}$. Additionally, in our expected factor, we do not consider the difference between the definitions of $\xi$ which is significant for the Newtonian case here as $\epsilon_B$ is not close to zero. Moreover, the full extent of electrons radiating at $\nu_m$ corrections can't be completely captured with simple factors since most exponents in our expressions are $\pb$ dependent. The expected factors are meant to capture primarily the direction of the differences i.e., whether the result of our analysis is larger or smaller and why.

\begin{table}[h]
    \centering
    \begin{tabular}{c|r|l||r|l}
        \hline
                                      & Relativistic    &        & Newtonian       & \\
        \hline
                                      & Expected Factor & Factor & Expected Factor & Factor \\
        \hline
        $R_{\rm{ours}}/R_{\rm{BNP13}}$     & $0.1^{\frac{1}{2\pb+13}}(\frac{\curlc}{3})^\frac{\pb+5}{2\pb+13}\approx 1.07$ 
                                        & $1.24$ at $8.30\sigma$ 
                                        & $4^{-\frac{1}{2\pb+13}}0.1^{\frac{1}{2\pb+13}}(\frac{\curlc}{3})^\frac{\pb+5}{2\pb+13}\approx 1.00$ 
                                        & $1.09$ at $0.60\sigma$ \\
        $E_{\rm{ours}}/E_{\rm{BNP13}}$     & $0.1^{-\frac{2(\pb+1)}{2\pb+13}}(\frac{\curlc}{3})^\frac{3(\pb+1)}{2\pb+13}\approx 3.41$ 
                                        & $5.17$ at $6.54\sigma$ 
                                        & $4^{-\frac{11}{2\pb+13}}0.1^{-\frac{2(\pb+1)}{2\pb+13}}(\frac{\curlc}{3})^\frac{3(\pb+1)}{2\pb+13}\approx 1.50$ 
                                        & $1.08$ at $0.46\sigma$ \\
        $B_{\rm{ours}}/B_{\rm{BNP13}}$     & $(\frac{3}{\curlc})^2(\frac{R_{\text{ours}}}{R_{\text{BNP13}}})^4\approx 0.93$ 
                                        & $0.78$ at $2.02\sigma$ 
                                        & $(\frac{3}{\curlc})^2(\frac{R_{\text{ours}}}{R_{\text{BNP13}}})^4 \approx 0.56$ 
                                        & $0.47$ at $10.32\sigma$ \\
        $N_{\rm{e,ours}}/N_{\rm{e,BNP13}}$ & $(\frac{\curlc}{3})^2(\frac{R_{\text{ours}}}{R_{\text{BNP13}}})^{-4}\approx 1.07$ 
                                        & $1.39$ at $3.55\sigma$
                                        & $4(\frac{\curlc}{3})^2(\frac{R_{\text{ours}}}{R_{\text{BNP13}}})^{-4}\approx 7.21$ 
                                        & $4.31$ at $5.29\sigma$ \\
   $n_{\rm{ext,ours}}/n_{\rm{ext,BNP13}}$  & $(\frac{12f_V}{f_{\Omega}})(\frac{N_{\text{e,ours}}}{N_{\text{e,BNP13}}})(\frac{R_{\text{ours}}}{R_{\rm{C24}}})^{-3}\approx 8.75$ 
                                        & $9.35$ at $3.62\sigma$ 
                                        & $(\frac{12f_V}{f_{\Omega}})(\frac{N_{\text{e,ours}}}{N_{\text{e,BNP13}}})(\frac{R_{\text{ours}}}{R_{\text{BNP13}}})^{-3}\approx 3.60$ 
                                        & $3.81$ at $3.59\sigma$ \\
        \hline
    \end{tabular}
    \caption{Expected factors considering dominant corrections compared against the ratio of our equipartition analysis values and that of \cite{Christy_et_al._2024} which utilizes the formalism of \cite{Barniol_Duran_Nakar_Piran_2013} for the TDE ASASSN-19bt (Section~\ref{ASASSN-19btSubsectionLabel}). The ratio and its standard deviations $\sigma$ away from agreement with unity are time averaged. For the off-axis relativistic case both off-axis values are averaged together. Powers of $\curlc/3$ are $\curlc\neq3$ corrections. Powers of $0.1$ are the hot proton affecting DFE corrections. Powers of $4$ are the \cite{Barniol_Duran_Nakar_Piran_2013} corrections that we neglect. The factor $12f_V/f_\Omega$ reflects the differences in method for the derivation of $n_e$ between \cite{Matsumoto_Piran_2023} and \cite{Christy_et_al._2024}.}
    \label{significancetable}
\end{table}

For the off-axis relativistic cases, we find that the estimated $\gamma_m$ is greater than the peak Lorentz factor $\gamma_e=\gamma_a$, inconsistent with the assumed ordering $\nu_m<\nu_a$, and we set $\kappa=1$ to account for this. For the Newtonian case, however, we find $\gamma_m<\gamma_e=\gamma_a$ (we set $\gamma_m=2$ in this case) consistent with the assumed spectral ordering. This suggests that, when $\nu_m<\nu_a$ is assumed, the off-axis relativistic cases might be unphysical for ASASSN-19bt and the Newtonian interpretation should be preferred. Here we have not considered an off-axis relativistic case with assumed spectral ordering $\nu_a<\nu_m$, leaving such exploration to future work.

In summary, we have found that the radius and energy predicted in our equipartition framework for ASSASN-19bt in the off axis relativistic cases is about $5$ times larger when additionally considering the interdependence of the various corrections. The differences between our results and that of \citep{Christy_et_al._2024} are largely explained by our more complete treatment of $\curlc$ and our significantly smaller definition for the out of equipartition parameter $\epsilon$. In the Newtonian case we have not found a significantly larger energy but this is the result of the various corrections coincidentally cancelling out rather than a generally applicable result.

\subsection{AT2019dsg}
\begin{figure}[h]
    \centering
    \includegraphics[width=1\linewidth]{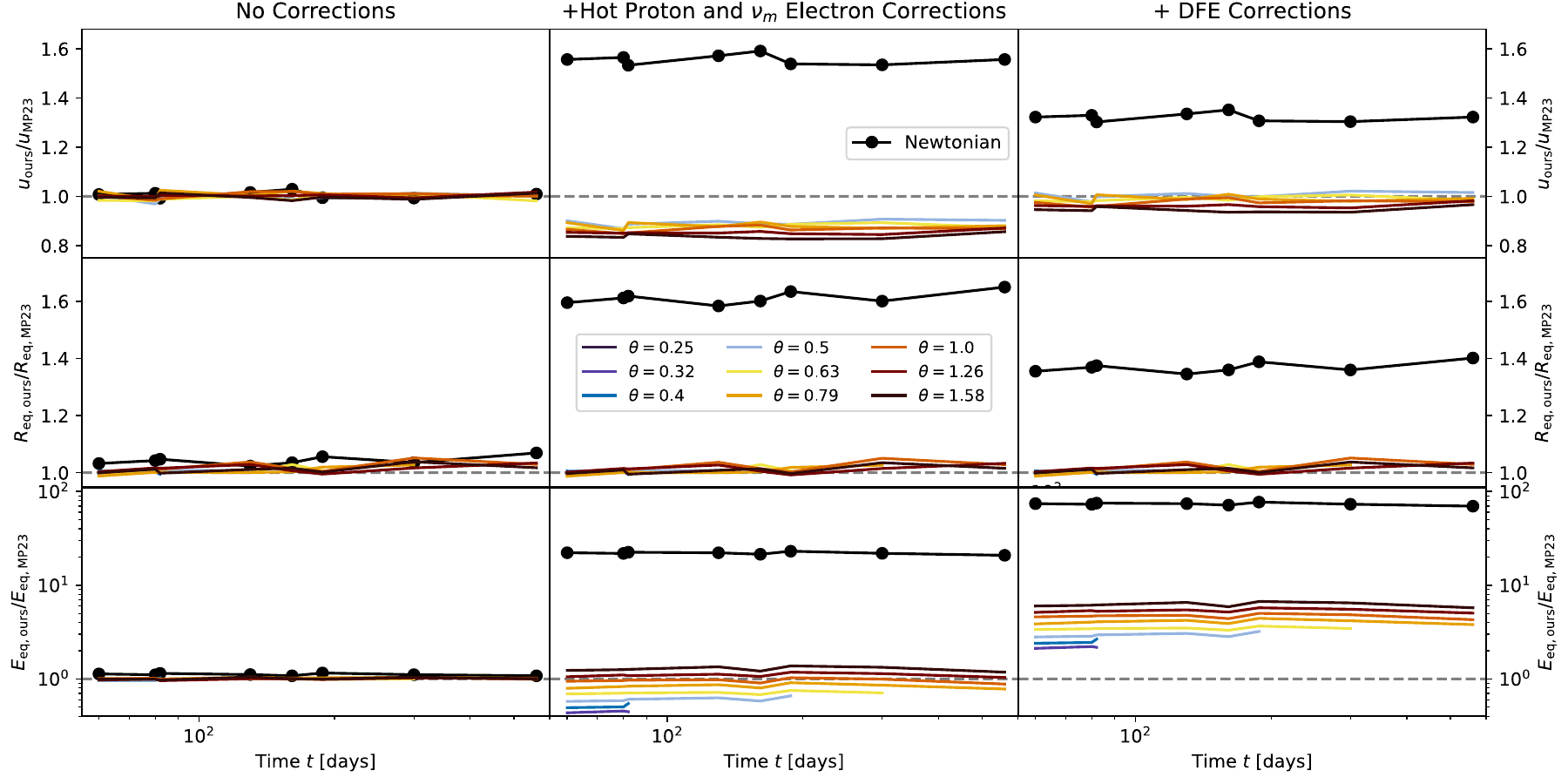}
    \caption{The ratio of equipartition quantities derived from our analysis of AT2019dsg with that using the framework of \cite{Matsumoto_Piran_2023} for a Newtonian case (black) and off-axis case (colored) for no corrections $\curlc=3$, $\xi=1$, $\pb=2$, $\epsilon=1$ (left), for only hot proton and electrons radiating at $\nu_m$ corrections $\curlc\neq3$, $\xi\neq1$, $\pb\neq2$, $\epsilon=1$ (middle), and for deviations from equipartition $\curlc\neq3$, $\xi\neq1$, $\pb\neq2$, $\epsilon\neq1$ (right). Here we assume the same cosmology as \cite{Matsumoto_Piran_2023} $H_0=69.6\text{ km/s/Mpc}$, $\Omega_{m}=0.286$, $\Omega_{\Lambda}=0.714$, the same geometric parameters as \cite{Matsumoto_Piran_2023} $f_A=f_V=1$, and the same microphysical parameters $\epsilon_e=0.1$, $\epsilon_B=0.02$ and redshift $z=0.051$ in \cite{Cendes_Alexander_Berger_Eftekhari_Williams_Chornock_2021}.
    }
    \label{MP23comparisonFigure}
\end{figure}

Using our framework, we next perform an analysis of AT2019dsg using the peak flux and peak frequency values in Table 2 of \cite{Cendes_Alexander_Berger_Eftekhari_Williams_Chornock_2021} and compare our analysis with that of \cite{Matsumoto_Piran_2023} in Figure \ref{MP23comparisonFigure}. Since \cite{Matsumoto_Piran_2023} do not provide tables of their derived parameters, we extract their values for 4-velocity, equipartition radius, and equipartition energy from their figures using WebPlotDigitizer. We first ensure that our code reproduces their analysis, which does not include $\curlc$, electrons radiating at $\nu_m$, hot proton, or DFE corrections\footnote{There is a small (factor of $\lesssim 1.2$) but negligible systematic difference between the results as we take, in the Newtonian case, $\delta_D=1$ whereas \cite{Matsumoto_Piran_2023} takes $\delta_D=(1-\beta)^{-1}$}. 
Since we obtained their analysis from their from plots, for some angles of the off-axis solutions the data was cut off by the limits of the plot. We then consider the corrections of electrons radiating at $\nu_m$, hot protons, and a full consideration of $\curlc\approx 4.565$ for $p=2.7$, but exclude DFE to examine such corrections in equipartition. We then, additionally, consider DFE for comparison in Figure \ref{MP23comparisonFigure}. Our new results, including DFE, are shown in Figure \ref{MP23corrREfigure}. Following \cite{Matsumoto_Piran_2023} we do not consider the \cite{Barniol_Duran_Nakar_Piran_2013} factors of 4.

With all corrections turned off in our model we recover the results of \cite{Matsumoto_Piran_2023} for the 4-velocity, equipartition radius, and equipartition energy for both the off-axis relativistic and Newtonian cases. For the additional considerations of only hot proton and electrons radiating at $\nu_m$ corrections we find an equipartition radius that is greater than the \cite{Matsumoto_Piran_2023} estimate in the Newtonian case by a factor of $1.61$, whereas there is agreement with the previous analysis for the off-axis relativistic case. Since for $\beta$, the apparent velocity, $u\approx \beta\propto R$ in the Newtonian case, the 4-velocity in this case is increased by about the same factor as the radius. In the off-axis case, the 4-velocity estimate is $1.14$ times smaller (for $\theta=0.79$). This decrease is greater for larger viewing angles. The energy in the Newtonian case is increased by a factor of $\sim 22$ whereas in the off-axis relativistic case it is increased slightly for large angles ($\times1.27$ for $\theta=1.58$) and decreased for small angles ($\times0.45$ for $\theta=0.32$). 

Electrons radiating at $\nu_m$ decrease our estimate of the radius by $\gamma_m^{\frac{2-\pb}{2\pb+13}}\approx 0.97$ in the Newtonian case ($\gamma_m=2$) and $\approx 0.793$ in the off-axis relativistic case ($\gamma_m\approx 450$)\footnote{There are additional corrections in the exponents of all the terms from consideration of electrons radiating at $\nu_m$, which we do not discuss here.}. Hot protons increase our estimate of the radius negligibly by  $\xi^\frac{1}{2\pb+13}\approx 1.10$ (for $\epsilon_e=0.1$ and the in equipartition case $\epsilon_B=\frac{2(\pb+1)}{2\pb+13}\approx 0.402$). Our full consideration of $\curlc$ increases our estimate of the radius in both cases by $(\curlc/3)^\frac{\pb+5}{2\pb+13}\approx 1.19$. Thus, the much higher $\gamma_m$ in the off-axis relativistic case causes a greater decrease from electrons radiating at $\nu_m$ to the radius that essentially cancels the hot proton and $\curlc$ corrections. Since the 4-velocity in the Newtonian case is proportional to the radius ($\beta_{\rm{\rm{eq,N}}}^{+\rm{dev}}\propto R_{\rm{\rm{eq,N}}}^{+\rm{dev}}$), the greater value found in our analysis can be explained from the differences in our radius estimate. For the off-axis relativistic case this greater Newtonian velocity decreases the critical angle $\theta_c$ which decreases our estimate of the off-axis Lorentz factor as the constraint function $f$ (see equation \ref{fourvelConstraint}) increases. Similarly, $f$ increases with $\theta$ so that off-axis solutions with larger observer angles have a higher Lorentz factor. The contributions to the equipartition energy of electrons radiating at $\nu_m$ (factor of $\gamma_m^{\frac{11(2-\pb)}{2\pb+13}}\approx0.078$), hot protons (a factor of $\xi^\frac{11}{2\pb+13}\approx2.85$) and taking $\curlc\not=3$ (a factor of $(\curlc/3)^\frac{3(\pb+1)}{2\pb+13}\approx 1.29$) for the off-axis relativistic case are nearly cancelled out by the contributions of the relativistic corrections (a factor of $\Gamma\delta^{-\frac{7\pb+29}{2\pb+13}}\propto2\Gamma^{-\frac{5\pb+16}{2\pb+13}}$ for small angles) since we find a smaller Lorentz factor for larger angles.

We then consider the additional effect of adding DFE corrections. The magnetic energy density fraction is small $\epsilon_B=0.02$ in this case so that $\epsilon\ll1$. The 4-velocity and equipartition radius both decrease for the Newtonian case since $\beta_{\rm{\rm{eq,N}}}^{+\rm{dev}}\propto R_{\rm{\rm{eq,N}}}^{+\rm{dev}}\propto\epsilon^{1/(2\pb+13)}\approx 0.83$. The equipartition radius in the relativistic case is largely unaffected by DFE since the 4-velocity in the relativistic case is closer to the no-corrections estimate than from the decreased estimate with only hot proton and electrons radiating at $\nu_m$ corrections. The addition of DFE brings the relativistic 4-velocity closer to the no corrections estimate since the (approximate) critical angle $\thetactilde\propto R_{\rm{eq,N+dev}}^{-\frac{2\pb+13}{3(\pb+6)}}\propto\epsilon^{-\frac{1}{3(\pb+6)}}$ which increases the off-axis estimate. The equipartition energy in both off-axis relativistic and Newtonian cases is increased as the factor $(11/17)\epsilon^{-2(\pb+1)/(2\pb+13)}\approx 2.64$ dominates for small $\epsilon$.

\begin{figure}[!htb]
    \centering
    \includegraphics[width=0.8\linewidth]{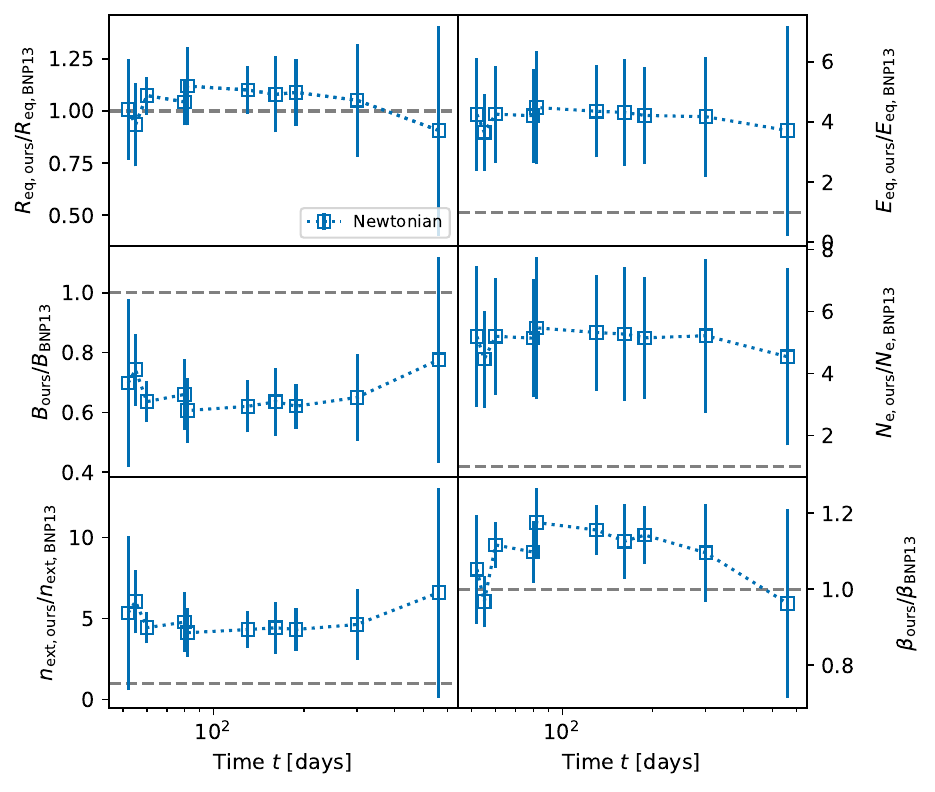}
    \caption{The ratio between the equipartition quantities derived from our analysis of AT2019dsg and that of \cite{Cendes_Alexander_Berger_Eftekhari_Williams_Chornock_2021}, which uses the \cite{Barniol_Duran_Nakar_Piran_2013} formalism, for a Newtonian model. Here we assume the same geometry as \cite{Cendes_Alexander_Berger_Eftekhari_Williams_Chornock_2021} $f_V=0.36$, $f_A=1$, and same microphysical parameters $\epsilon_e=0.1$, $\epsilon_B=0.02$ and the \cite{Planck_2018} Table 2 (TT, TE, EE + lowE + lensing + BAO) cosmology $H_0=67.66\text{ km/s/Mpc}$, $\Omega_{m}=0.3111$, $\Omega_{\Lambda}=0.6889$.}
    \label{CendesCompFigureLabel}
\end{figure}

We also compare our results with that of \cite{Cendes_Alexander_Berger_Eftekhari_Williams_Chornock_2021} (see Figure \ref{CendesCompFigureLabel}) which, in the same fashion as \cite{Christy_et_al._2024}, applied the individual corrections of \cite{Barniol_Duran_Nakar_Piran_2013} together. We find a significantly larger energy $E$, electron number $N_e$, and number density $n_{\rm{ext}}$, and a significantly smaller magnetic field. These differences are the same effects as discussed for the Newtonian case in section \ref{ASASSN-19btSubsectionLabel}. The energy in our analysis of ASASSN-19bt is increased more in the Newtonian case than for AT2019dsg due to the smaller $\epsilon_B$ for ASASSN-19bt.

In summary, we have performed an equipartition analysis of AT\,2019dsg using our framework including hot proton, electrons radiating at $\nu_m$, DFE, and $\curlc$ corrections considering a Newtonian and an off axis model. When all corrections are self-consistently considered, the energy is increased by a factor $\sim70$ for the Newtonian case and by a factor $\sim4$ for the off axis case ($\theta=0.79$) from the \citep{Matsumoto_Piran_2023} estimate. When we turned off the corrections, we were able to recover the results of \citep{Matsumoto_Piran_2023}. Our comparison to the results of \cite{cendes_ubiquitous_2023} for the Newtonian case shows a significant portion of this increase (a factor of $\sim4$) in energy is caused primarily by the self-consistent treatment of hot proton and DFE corrections.

\subsection{EP240414a}

Using the framework of \cite{Matsumoto_Piran_2023}, \cite{Bright_etal_2025} performed an equipartition analysis of the fast X-ray transient EP240414a ($z=0.4018\pm0.0010$) assuming an on-axis jet with an SSA peak. We similarly perform an equipartition analysis of EP240414a for an on-axis jet, additionally considering electrons radiating at $\nu_m$, hot protons, and the full expression for $\curlc$. As in \cite{Bright_etal_2025}, we assume the emission peaks at $t_{obs}=30\text{ days}$ with $F_p=434\pm21\text{ $\mu$Jy}$, $\nu_p=3\text{ GHz}$, $p=2.16\pm0.3$, and that the system is in equipartition $\epsilon$, since neither $\nu_c$ nor $R$ was observationally constrained. We propagate uncertainties assuming that inputs to the equipartition analysis are normally distributed with no covariance. The uncertainty is significantly and systematically underestimated due to the assumption (made because of sparse temporal sampling) that the emission peaks strictly at $\nu_p=3\text{ GHz}$ at 30 days. However, \cite{Bright_etal_2025} show that the choice of $\nu_p=3\text{ GHz}$ yields a conservative estimate on the apparent velocity $\beta_{\rm{\rm{eq,N}}}$ and still implies a mildly relativistic outflow. This same reasoning holds for the corrections we consider, since they only have a small effect on the Newtonian equipartition radius $R_{\rm{\rm{eq,N}}}$ and $\beta_{\rm{\rm{eq,N}}}=(1+z)R_{\rm{\rm{eq,N}}}/(ct)$. Like \cite{Bright_etal_2025}, we also solve for the Lorentz factor numerically. We assume the same cosmology ($H_0=70\text{ km/s/Mpc}$, $\Omega_{m}=0.3$, $T_{CMB}=2.725\text{ K}$), observer angle $\theta_{obs}=0$, and geometric parameters $f_A=f_V=f_\Omega=1$. We also assume typical $\epsilon_e=0.1$.

The equipartition energy we find for EP240414a, $E_{\rm{eq}}=(6.3\pm0.3)\times10^{48}\text{ erg}$, is $\sim 6$ times larger than the value found by \cite{Bright_etal_2025} of $E_{\rm{eq}}\sim 10^{48}$. This suggests a higher minimum energy for the object for typical $\epsilon_e=0.1$. The Lorentz factor we find $\Gamma\approx1.71\pm0.04$ is comparable to the value  of $\Gamma=1.6$ found by \cite{Bright_etal_2025}. We also find $R_{\rm{eq}}=(2.36\pm0.16)\times10^{17}\text{ cm}$, $B=(6.47\pm0.25)\times10^{-2}\text{ G}$, $N_e=3.06\times10^{52}$ with 75\% CI $[8.81\times10^{51}, 5.62\times10^{52}]$, and $n_\text{ext}=7.33\text{ cm$^{-3}$}$ with 75\% CI $[1.54, 14.24]\text{ cm$^{-3}$}$. We quote confidence intervals for the number of swept up electrons and the ambient number density due to their wide distribution. This wide distribution is due to the fact that $N_e$ and $n_\text{ext}\propto N_e$ are sensitive to electrons radiating at $\nu_m$ corrections as $N_e\propto(\gamma_e/\gamma_m)^{1-p}$ and there is a large uncertainty in $p$.

In summary, we performed an in equipartition analysis of EP240414a using our framework considering electrons radiating at $\nu_m$, hot proton (assuming $\epsilon_e=0.1$), and $\curlc\approx3.42$ (for $p=2.16$) corrections. We find an energy that is increased by a factor $\sim 6$ for $\epsilon_e=0.1$. Most of the increase in the energy can be accounted for from hot proton and electrons radiating at $\nu_m$ corrections alone as the factor $(\curlc/3)^\frac{2(\pb+1)}{2\pb+13}\approx 1.1$ is close to unity in this case. 

\subsection{J0231-0433}
We perform an equipartition analysis of AGN J0231-0433 ($z=0.188$) incorporating hot proton and electrons radiating at $\nu_m$ corrections. We assume the same peak flux $F_p=330\pm60\text{ mJy}$, peak frequency $\nu_{p}=272\pm36\text{ MHz}$, and a $\nu_a$ peaked spectrum with optically thin spectral index $\alpha_\text{thin}=-0.57\pm0.16=(1-p)/2$ as Table 1 of \cite{Keim_etal_2019}. When measuring the angular diameter of J0231-0433, \citep{Keim_etal_2019} identifies an eastern and western component of emission. Where applicable, we use the properties of the brighter western component. We also assume the same cosmology: $H_0=70\text{ km/s/Mpc}$, $\Omega_{m}=0.28$, $\Omega_{\Lambda}=0.72$. We additionally assume a Newtonian outflow in equipartition with typical $\epsilon_e=0.1$ and geometry $f_A=1$, $f_V=0.36$, $f_\Omega=4$. Since the synchrotron emission of AGN varies much more slowly than the other objects considered here, we assume a static emitting region by taking the limit $t\to\infty$ (so that $\gamma_m=2$ at the minimum) and ignore synchrotron cooling. In this context, the radius of the shock $R$ becomes the size of the emitting region. We propagate errors assuming that model parameters are normally distributed with no covariance. 

From our equipartition analysis we find a magnetic field strength of $B=9.17\pm1.99\text{ mG}$, an energy of $E_{\rm{eq}}=(5.13\pm3.96)\times10^{53}\text{ ergs}$ and a radius for the emitting region of $R_{\rm{eq}}=(3.56\pm0.694)\times10^{19}\text{ cm}$. We also find $N_e=(3.08\pm2.37)\times10^{58}$, $n_\text{ext}=0.15\pm0.06\text{ cm$^{-3}$}$. \cite{Keim_etal_2019} estimated the magnetic field of AGN J0231-0433 by constraining the size of the emitting region $R$ using measurements of the major and minor axes angular sizes instead of assuming equipartition.  
\cite{Keim_etal_2019} find a magnetic field about an order of magnitude less of $B=0.5\pm0.4 \text{ mG}$ for the brighter, western, component. This suggests that the equipartition radius overestimates the actual radius of the emitting region so that an out of equipartition analysis with $\epsilon_B\ll 1$ might be preferred since $B(R(\epsilon))\propto R(\epsilon)^4\propto \epsilon^\frac{4}{2\pb+13}\approx \epsilon_B^\frac{4}{2\pb+13}$. 

\begin{figure}[b]
    \centering
    \includegraphics[width=1\linewidth]{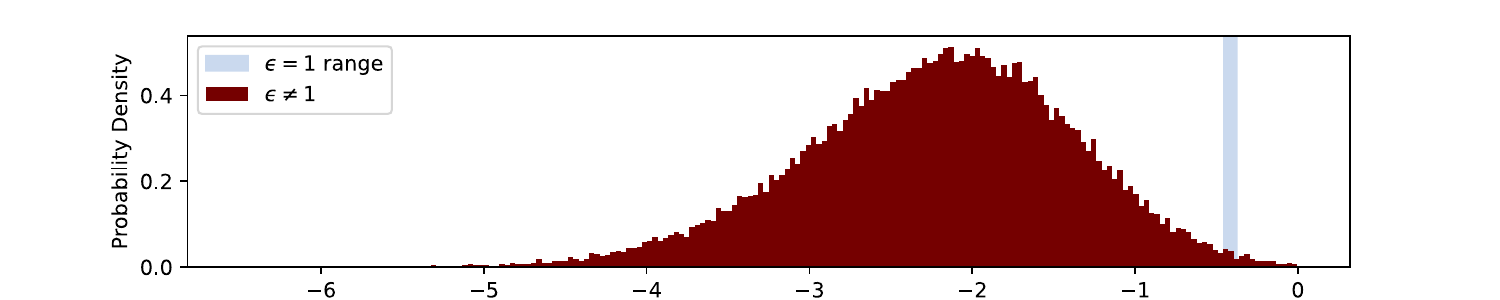}
    \caption{The probability density of $\epsilon_B$ assuming that the \cite{Keim_etal_2019} estimate for the radius of the emitting region $R=(2.843\pm0.302)\times10^{19}\text{ cm}$ of J0231-0433 is normally distributed. The possible values of $\epsilon_B$ for an in equipartition $\epsilon=1$ scenario are for $2<p<3$. There is a maximum $\sim0.3$\% chance the system is in equipartition when considering a DFE $\epsilon\neq1$ scenario.}
    \label{epsBfigure}
\end{figure}

\cite{Keim_etal_2019} (their Table 2) estimate a (geometric) average angular diameter of the brighter, western, component as $\theta_D=5.85\pm0.62\text{ mas}$
which corresponds to an emitting region radius of $R=(2.843\pm0.302)\times10^{19}\text{ cm}$. Assuming that $\theta_D$ is normally distributed, we numerically solve for $\epsilon_B$ by equating $R=R_{\rm{eq}}^{\rm{+dev}}$ and perform and an equipartition analysis for the brighter component of J0231-0433 including DFE. We find a mean $\epsilon_B=0.0281$ with a 50\% CI of $[0.0036, 0.049]$ where $\log(\epsilon_B)$ is approximately normally distributed (see Figure \ref{epsBfigure}). We find an energy $E_{\rm{eq}}^{+\rm{dev}}=3.42\times10^{54}\text{ erg}$ with 50\% CI $[1.25\times10^{54}, 5.58\times10^{54}] \text{ erg}$, number of emitting electrons $N_e= 2.05\times10^{59}$ with 50\% CI $[7.54\times10^{58}, 3.35\times10^{59}]$, and ambient density $n_\text{ext}=2.69\text{ cm$^{-3}$}$ with 50\% CI $[0.85,6.41]\text{ cm$^{-3}$}$. As expected, the magnetic field strength $B=3.9\pm1.9\text{ mG}$ is smaller than the in-equipartition estimate and we recover the radius $R_{\rm{eq}}^{+\rm{dev}}=(2.836\pm0.635)\times10^{19}\text{ cm}$ of \cite{Keim_etal_2019}. We do not recover the same magnetic field as \citet{Keim_etal_2019} due to differences in the numerical prefactor between our expressions\footnote{The precise numerical prefactors used in equation 3 of \cite{Keim_etal_2019} are discussed in \cite{Gould_1979} and \cite{Marscher_1983}.}.

In summary, we performed an equipartition analysis for the brighter, western, component of J0231-0433 considering electrons radiating at $\nu_m$, hot proton, DFE, and $\curlc$ corrections. We constrained a value for $\epsilon_B$ using the measured radius from \cite{Keim_etal_2019}. The value found for $\epsilon_B$ is small implying that consideration of the effects of hot protons on DFE corrections significantly contributes to the factor $\sim7$ increase in the energy from the minimum in equipartition estimate. Even when considering DFE, we find a magnetic field that is about an order of magnitude larger.

\section{Conclusion}
We have developed a self-consistent equipartition framework for Newtonian, on-axis relativistic, and off-axis relativistic synchrotron outflows that simultaneously accounts for hot protons, electrons radiating at $\nu_m$, deviations from equipartition, and the full expression for the constant $\curlc$ needed to match the SSA and Rayleigh-Jeans flux. We find that the minimum energy state corresponds to $\epsilon_B\approx0.3$, implying that $\epsilon_B$ alone determines whether a system is in equipartition. In cases when both $\nu_a$ and $\nu_m$ can be identified, we have derived additional expressions for the equipartition energy and radius so that $\gamma_m$ need not be theoretically estimated. 
Our expressions recover previous works  \citep{Matsumoto_Piran_2023,Sfaradi_et_al_2025,Cowie_Fender_2026} in the relevant limits. We have implemented our results in a freely available, open-source code. Applying our framework to the TDEs AT2019dsg and ASASSN-19bt, the fast X-ray transient EP240414a, and the AGN J0231-0433, we recover previous analyses when all correction factors are disabled, validating our implementation. Applying the full self-consistent framework changes inferred physical parameters by factors of a few. For the TDEs, the inferred energies and ambient densities increase by approximately a factor of 5, EP240414a yields an energy larger by roughly a factor of 6, and the inferred magnetic field of J0231-0433 increases by nearly an order of magnitude.

Our results demonstrate that treating equipartition corrections independently can introduce systematic biases into inferred source properties. While these differences are generally not large enough to alter the qualitative conclusions of previous studies, they are significant at the level of modern observational precision. In particular, the equipartition energies of sources with typical microphysical parameters may be systematically underestimated by factors of several if the interdependence of the correction factors is neglected. We excluded the \citep{Barniol_Duran_Nakar_Piran_2013} factors of 4 from our analyses, which represent small geometrical corrections in the Newtonian regime. Their applicability is less certain for for mildly relativistic cases. Extending these geometric corrections to provide a smooth description across both regimes is a natural next step and will be the focus of future work. Together with the self-consistent framework presented here, such extensions will provide a robust foundation for equipartition analyses across the full range of synchrotron sources, from Newtonian outflows to relativistic jets.

\section{Acknowledgments}
We thank Dr. Tatsuya Matsumoto for insightful discussions. N.F.\ acknowledges support from the National Science Foundation Graduate Research Fellowship Program under Grant No. DGE-2137419.

\section{Appendix A}
\begin{figure}[h!]
    \centering
    \includegraphics[width=0.8\linewidth]{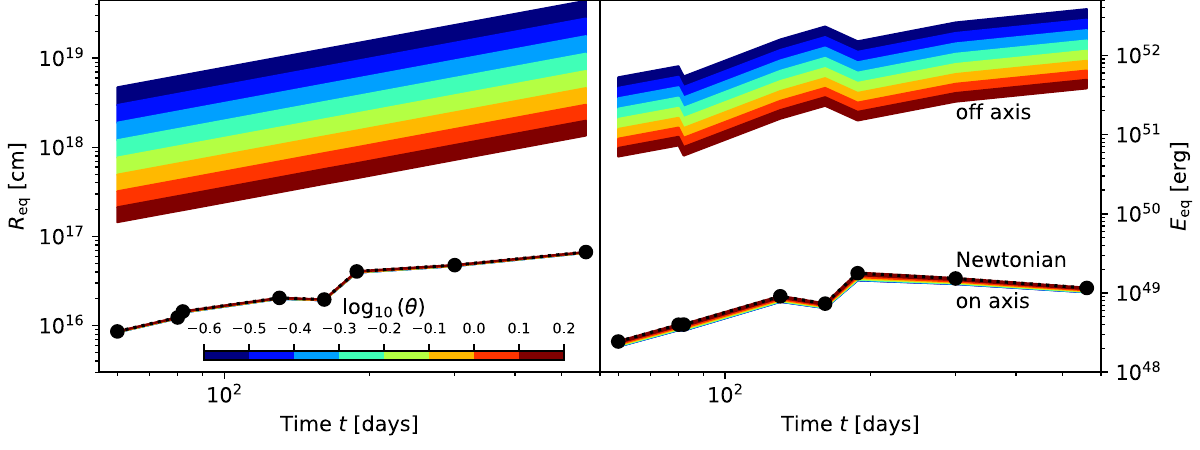}
    \caption{The equipartition radius and energy for the Newtonian, on axis, and off-axis cases for AT2019dsg including all corrections $\curlc\neq3$, $\xi\neq1$, $\pb\neq2$, $\epsilon\neq1$. Here we assume the same cosmology as \cite{Matsumoto_Piran_2023} $H_0=69.6\text{ km/s/Mpc}$, $\Omega_{m,0}=0.286$, $\Omega_{0}=0.714$ and the same geometric parameters $f_A=f_V=1$. For the on axis solution we set $\gamma_m=2$ as the $\gamma_m$ we calculate from the bulk LF is small $<2$.}
    \label{MP23corrREfigure}
\end{figure}

Here we provide tables and graphs for the purpose of comparison in future works. In Tables \ref{ChristyReproducedTable} and \ref{CendesReproducedTable} we provide the results of our equipartition analysis for ASASSN-19bt and AT2019dsg, respectively. In Figure \ref{MP23corrREfigure} we plot our results for AT2019dsg for various observer angles. 

\begin{table}[h!]
\centering
\begin{tabular}{l|ccccccc}
Spherical & $t$ & $\log(R)$ (cm) & $\log(E)$ (ergs) & $\log(B)$ (G) & $\log(N_e)$ & $\log(n_{\text{ext}})$ ($\text{cm}^{-3}$) & $\beta$\\
\hline
& $49$ & $15.33\pm0.06$ & $46.42\pm0.07$ & $0.53\pm0.05$ & $51.19\pm0.07$ & $4.57\pm0.10$ & $0.018\pm0.002$ \\
$f_A = 1.00$ & $88$ & $15.94\pm0.07$ & $47.10\pm0.08$ & $-0.04\pm0.07$ & $51.88\pm0.08$ & $3.42\pm0.14$ & $0.040\pm0.007$ \\
$f_V = 0.36$ & $132$ & $16.36\pm0.07$ & $47.75\pm0.09$ & $-0.34\pm0.07$ & $52.53\pm0.09$ & $2.83\pm0.13$ & $0.070\pm0.012$ \\
$\epsilon_e = 0.100$ & $186$ & $16.67\pm0.08$ & $48.19\pm0.11$ & $-0.59\pm0.07$ & $52.97\pm0.11$ & $2.33\pm0.15$ & $0.101\pm0.020$ \\
$\epsilon_B = 0.200$ & $457$ & $16.84\pm0.01$ & $48.95\pm0.02$ & $-0.46\pm0.01$ & $53.72\pm0.02$ & $2.58\pm0.03$ & $0.060\pm0.002$ \\
 & $800$ & $17.13\pm0.02$ & $49.21\pm0.03$ & $-0.77\pm0.02$ & $53.98\pm0.03$ & $1.96\pm0.04$ & $0.068\pm0.003$ \\
 & $1377$ & $17.45\pm0.02$ & $49.61\pm0.03$ & $-1.05\pm0.02$ & $54.38\pm0.03$ & $1.40\pm0.04$ & $0.082\pm0.004$ \\
\hline
Jet & $t$ & $\log(R)$ (cm) & $\log(E)$ (ergs) & $\log(B)$ (G) & $\log(N_e)$ & $\log(n_{\text{ext}})$ ($\text{cm}^{-3}$) & $\Gamma$\\
\hline
 & $49$ & $17.39\pm0.00$ & $51.22\pm0.07$ & $0.44\pm0.07$ & $52.98\pm0.04$ & $4.24\pm0.12$ & $53.414\pm5.037$ \\
$\theta_{\text{obs}}=$$1.05$ & $88$ & $17.64\pm0.00$ & $51.09\pm0.09$ & $-0.27\pm0.09$ & $53.00\pm0.06$ & $2.98\pm0.16$ & $29.383\pm3.423$ \\
$f_A = 1.00$ & $132$ & $17.82\pm0.00$ & $51.20\pm0.08$ & $-0.65\pm0.09$ & $53.19\pm0.06$ & $2.29\pm0.15$ & $19.678\pm2.281$ \\
$f_V = 1.00$ & $186$ & $17.97\pm0.00$ & $51.27\pm0.09$ & $-0.96\pm0.10$ & $53.31\pm0.06$ & $1.73\pm0.17$ & $14.989\pm1.932$ \\
$\epsilon_e = 0.100$ & $457$ & $18.36\pm0.00$ & $52.53\pm0.02$ & $-0.75\pm0.02$ & $54.39\pm0.01$ & $1.96\pm0.03$ & $21.517\pm0.512$ \\
$\epsilon_B = 0.003$ & $800$ & $18.60\pm0.00$ & $52.67\pm0.03$ & $-1.08\pm0.03$ & $54.57\pm0.02$ & $1.33\pm0.05$ & $19.784\pm0.712$ \\
 & $1377$ & $18.84\pm0.00$ & $52.88\pm0.03$ & $-1.39\pm0.03$ & $54.81\pm0.02$ & $0.74\pm0.05$ & $17.206\pm0.627$ \\
\hline
Jet & $t$ & $\log(R)$ (cm) & $\log(E)$ (ergs) & $\log(B)$ (G) & $\log(N_e)$ & $\log(n_{\text{ext}})$ ($\text{cm}^{-3}$) & $\Gamma$\\
\hline
 & $49$ & $17.09\pm0.00$ & $50.92\pm0.07$ & $0.53\pm0.07$ & $52.83\pm0.04$ & $4.35\pm0.12$ & $25.741\pm2.355$ \\
$\theta_{\text{obs}}=$$1.57$ & $88$ & $17.35\pm0.00$ & $50.79\pm0.09$ & $-0.18\pm0.10$ & $52.85\pm0.06$ & $3.09\pm0.16$ & $14.136\pm1.685$ \\
$f_A = 1.00$ & $132$ & $17.52\pm0.00$ & $50.90\pm0.08$ & $-0.56\pm0.09$ & $53.04\pm0.06$ & $2.40\pm0.15$ & $9.460\pm1.102$ \\
$f_V = 1.00$ & $186$ & $17.67\pm0.00$ & $50.97\pm0.09$ & $-0.86\pm0.10$ & $53.15\pm0.07$ & $1.84\pm0.17$ & $7.186\pm0.938$ \\
$\epsilon_e = 0.100$ & $457$ & $18.06\pm0.00$ & $52.23\pm0.02$ & $-0.66\pm0.02$ & $54.24\pm0.01$ & $2.07\pm0.03$ & $10.348\pm0.244$ \\
$\epsilon_B = 0.005$ & $800$ & $18.30\pm0.00$ & $52.37\pm0.03$ & $-0.99\pm0.03$ & $54.41\pm0.02$ & $1.44\pm0.05$ & $9.504\pm0.339$ \\
 & $1377$ & $18.54\pm0.00$ & $52.59\pm0.03$ & $-1.30\pm0.03$ & $54.65\pm0.02$ & $0.85\pm0.05$ & $8.276\pm0.304$ \\
\end{tabular}
\caption{Table of values for ASASSN-19bt including all corrections. Here we assume the \cite{Planck_2018} Table 2 (TT, TE, EE + lowE + lensing + BAO) cosmology $H_0=67.66\text{ km/s/Mpc}$, $\Omega_{m}=0.3111$, $\Omega_{\Lambda}=0.6889$, and the same microphysical and geometric parameters in each case as in \cite{Christy_et_al._2024}.}
\label{ChristyReproducedTable}
\end{table}

\begin{table}[h!]
\centering
\begin{tabular}{l|ccccccc}
Spherical & $t$ & $\log(R)$ (cm) & $\log(E)$ (ergs) & $\log(B)$ (G) & $\log(N_e)$ & $\log(n_{\text{ext}})$ ($\text{cm}^{-3}$) & $\beta$\\
\hline
 & $52$ & $15.92\pm0.03$ & $48.26\pm0.13$ & $0.07\pm0.03$ & $53.04\pm0.13$ & $4.65\pm0.07$ & $0.065\pm0.005$ \\
$f_A = 1.00$ & $55$ & $15.81\pm0.02$ & $48.20\pm0.13$ & $0.21\pm0.04$ & $52.98\pm0.13$ & $4.93\pm0.07$ & $0.048\pm0.003$ \\
$f_V = 0.36$ & $60$ & $15.97\pm0.02$ & $48.38\pm0.13$ & $0.06\pm0.03$ & $53.16\pm0.13$ & $4.63\pm0.07$ & $0.063\pm0.003$ \\
$\epsilon_e = 0.100$ & $80$ & $16.11\pm0.02$ & $48.59\pm0.13$ & $-0.04\pm0.03$ & $53.37\pm0.13$ & $4.42\pm0.07$ & $0.065\pm0.003$ \\
$\epsilon_B = 0.020$ & $82$ & $16.20\pm0.02$ & $48.61\pm0.13$ & $-0.17\pm0.03$ & $53.39\pm0.13$ & $4.17\pm0.07$ & $0.078\pm0.004$ \\
 & $130$ & $16.34\pm0.02$ & $48.96\pm0.13$ & $-0.21\pm0.03$ & $53.73\pm0.13$ & $4.09\pm0.07$ & $0.068\pm0.003$ \\
 & $161$ & $16.32\pm0.02$ & $48.86\pm0.13$ & $-0.23\pm0.03$ & $53.64\pm0.13$ & $4.05\pm0.07$ & $0.053\pm0.003$ \\
 & $188$ & $16.64\pm0.02$ & $49.25\pm0.13$ & $-0.51\pm0.03$ & $54.02\pm0.13$ & $3.49\pm0.07$ & $0.094\pm0.005$ \\
 & $300$ & $16.70\pm0.02$ & $49.18\pm0.13$ & $-0.64\pm0.03$ & $53.95\pm0.13$ & $3.23\pm0.07$ & $0.068\pm0.003$ \\
 & $561$ & $16.81\pm0.03$ & $49.03\pm0.13$ & $-0.87\pm0.03$ & $53.81\pm0.13$ & $2.76\pm0.07$ & $0.046\pm0.003$ \\
\end{tabular}
\caption{Table of values for AT2019dsg including all corrections. Here we assume the same geometry as \cite{Cendes_Alexander_Berger_Eftekhari_Williams_Chornock_2021}, and microphysical parameters and the \cite{Planck_2018} Table 2 (TT, TE, EE + lowE + lensing + BAO) cosmology $H_0=67.66\text{ km/s/Mpc}$, $\Omega_{m}=0.3111$, $\Omega_{\Lambda}=0.6889$.}
\label{CendesReproducedTable}
\end{table}

\clearpage

\bibliographystyle{aasjournalv7}
\bibliography{biblio}

\end{document}